\documentclass[aps,prl,superscriptaddress,notitlepage,amsmath,amsfonts,amssymb,citeautoscript,longbibliography,floatfix,twocolumn,10pt,reprint]{revtex4-2}

\usepackage{enumitem}
\usepackage{natbib}
\usepackage[english]{babel}
\usepackage{letltxmacro}
\usepackage{latexsym}

\usepackage{amsthm}
\usepackage{graphicx}
\usepackage{bm}
\usepackage{physics}
\usepackage[dvipsnames]{xcolor}
\usepackage{algorithm}
\usepackage{algpseudocode}
\usepackage{blindtext}
\usepackage{orcidlink}
\usepackage[normalem]{ulem}

\newcommand{\Ham}{\mathcal{H}}
\newcommand{\eff}{\text{eff}}
\newcommand{\Heff}{H_\eff}

\newcommand{\paradash}[1]{{#1.\textemdash}}
\makeatletter
\renewcommand{\paragraph}{%
	\@startsection
	{paragraph}%
	{4}%
	{\parindent}%
	{\z@}%
	{-0.2em}%
	{\normalfont\normalsize\itshape\bfseries\paradash}%
}%
\makeatother

\usepackage{hyperref}
\hypersetup{
	colorlinks=true,
	linkcolor=teal,
	citecolor=teal,
	filecolor=teal,
	urlcolor=teal,
	hypertexnames=false  
}

\begin{document}

\title{Ergodicity breaking in matrix-product-state effective Hamiltonians
}

\author{Andrew Hallam\textsuperscript{\orcidlink{0000-0003-2288-7661}}}
\thanks{These authors contributed equally.}
\affiliation{School of Physics and Astronomy, University of Leeds, Leeds LS2 9JT, UK}
\author{Jared Jeyaretnam\textsuperscript{\orcidlink{0000-0002-8316-9025}}}
\thanks{These authors contributed equally.}
\affiliation{School of Physics and Astronomy, University of Leeds, Leeds LS2 9JT, UK}
\affiliation{School of Physics and Astronomy, University of Nottingham, Nottingham, NG7 2RD, UK}
\affiliation{Centre for the Mathematics and Theoretical Physics of Quantum Non-Equilibrium Systems, University of Nottingham, Nottingham, NG7 2RD, UK}

\author{Zlatko Papi\'c\textsuperscript{\orcidlink{0000-0002-8451-2235}}}
\affiliation{School of Physics and Astronomy, University of Leeds, Leeds LS2 9JT, UK}

\begin{abstract}
Thermalization and its breakdown in interacting quantum many-body systems are governed by mid-spectrum eigenstates, which are typically accessible only in  small system sizes amenable to exact diagonalization.
Here we demonstrate that the density-matrix renormalization group (DMRG) effective Hamiltonian, an object routinely used to variationally approximate ground states,
encodes detailed information about the dynamics far from equilibrium.
In the random-field XXZ spin chain, the spectrum of the effective Hamiltonian is shown to capture the transition from thermal to many-body localized regimes, including spatially resolved probes of ergodic bubbles.
Furthermore, the same approach also captures weak ergodicity breaking associated with quantum many-body scars.
Our results establish the DMRG effective Hamiltonian as a versatile spectral probe of quantum thermalization and its breakdown in large systems beyond exact diagonalization.
\end{abstract}

\maketitle

\paragraph{Introduction}
Interacting quantum many-body systems exhibit an extraordinary richness of collective behavior, which is often challenging to characterize due to the exponential growth of their Hilbert spaces with system size.
In one spatial dimension (1D), matrix-product-state (MPS) methods~\cite{CiracRMP} and the related density-matrix renormalization group (DMRG)~\cite{White1992} have played a vital role in enabling controlled access to large system sizes.
These methods efficiently represent low-entanglement states, making it possible to accurately resolve ground state properties, excitation gaps, and critical behavior.

By contrast, extracting properties at \emph{finite} energy density above the ground state remains a major challenge.
In this regime, eigenstates typically obey the eigenstate thermalization hypothesis (ETH)~\cite{DeutschETH,SrednickiETH} and are highly entangled, restricting the numerics to small systems accessible by exact diagonalization (ED).
This part of the spectrum, however, is central to fundamental phenomena related to the breakdown of ergodicity, e.g.,  many-body localization (MBL)~\cite{Gornyi2005,Basko06,PalHuse}, where strong disorder fully suppresses thermalization; and quantum many-body scars (QMBSs)~\cite{Bernien2017,Turner2017}, which correspond to weakly-entangled eigenstates embedded within a thermal spectrum.
These phenomena are ubiquitous across many experimental platforms such as ultracold atoms and trapped ions~\cite{Schreiber2015,Smith2016, Rispoli2019,Jepsen2022,Leonard2023,yang2024phantom,GuoXian2022}, disordered semiconductors~\cite{Stanley2023}, Rydberg atom arrays~\cite{Bernien2017,  zhao2024observationquantumthermalizationrestricted, mark2025observationballisticplasmamemory,hudomal2025ergodicitybreakingmeetscriticality}, and superconducting circuits~\cite{Roushan2017,Guo2021,Yao2023,Dong2023,nagao2026probingmanybodylocalizationcrossover}.

\begin{figure}
    \includegraphics[width=0.99\columnwidth]{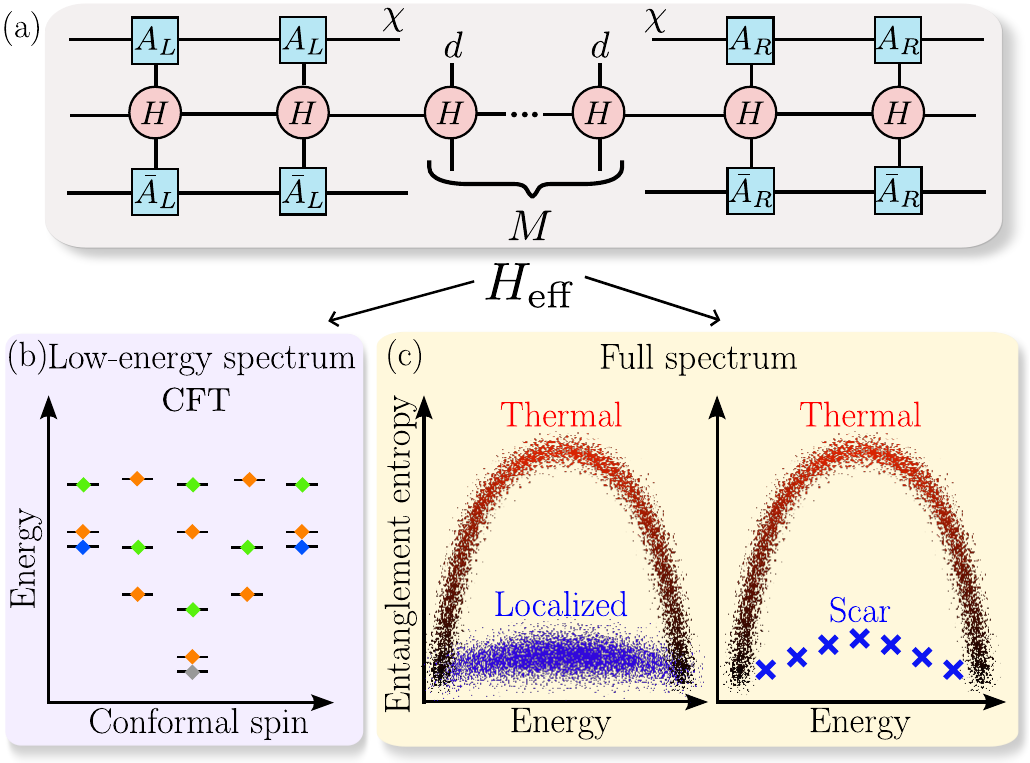}
    \centering
    \caption{%
	 	(a) The DMRG effective Hamiltonian $H_\mathrm{eff}$, supported on a subsystem of $M$ lattice sites.
		$H_\mathrm{eff}$ is constructed from the MPS left- and right-canonical tensors, $A_L$ and $A_R$, with maximum bond dimension $\chi$ and physical dimension $d$, and the MPO of the Hamiltonian, $\Ham$.
	 	(b) The low-lying spectrum of $H_\mathrm{eff}$ was shown to encode the conformal field theory (CFT) describing the critical point~\cite{Chepiga2017,Eberharter2023,cocchiarella2025excitedstateslocaleffective}.
	 	(c) Beyond low-lying excitations, in this work we show the full spectrum of $H_\mathrm{eff}$ captures ergodicity breaking phenomena, such as many-body localization and quantum many-body scars, which manifest as weakly-entangled nonthermal eigenstates.
    }
    \label{fig:summary}
\end{figure}

A natural question is: what role can MPS methods play in the study of ergodicity breaking phenomena?
Beyond simulations of time evolution~\cite{Znidaric08, Bardarson2012, Kloss2018, Doggen2018ManybodyLocalizationDelocalization, Chanda2020TimeDynamics, Sierant2022, Andraschko2014, Enss2017ManyBodyLocalizationInfiniteChains}, progress has been made in targeting individual excited eigenstates using MPS-based methods -- known as DMRG-X for MBL eigenstates~\cite{Khemani2016,Lim2016,Serbyn16E,Yu2017,Villalonga2018}, and DMRG-S for QMBS eigenstates~\cite{DMRGS} (see also the reviews~\cite{LuitzLevReview,AletReview,Sierant2025Jan}).
Furthermore, tensor networks have been used to approximate local integrals of motion in the MBL phase~\cite{Serbyn13-1,Huse13}, effectively diagonalizing the Hamiltonian at strong disorder~\cite{Pollmann2016,Pekker2017,Wahl2017,Wahl2019,Chertkov2021}.
While powerful in appropriate regimes, these approaches are intrinsically biased toward low entanglement and thus have not been expected to faithfully describe the thermal phase.
Alternative approaches based on matrix-product operators combined with polynomial expansions can probe spectral functions and densities of states, though without access to individual eigenstates~\cite{YilunYang2020}.
None of these methods, however, have been used to reconstruct level statistics or eigenvalue correlations, which are among the most sensitive diagnostics of thermalization.
On the other hand, it has been shown that the effective Hamiltonian -- a central object arising in the DMRG procedure, Fig.~\ref{fig:summary}(a) -- \emph{does} encode the low-energy spectra at quantum critical points, Fig.~\ref{fig:summary}(b)~\cite{Chepiga2017,Eberharter2023,cocchiarella2025excitedstateslocaleffective}.
However, these studies have focused on vanishing energy densities, leaving open the question of whether similar ideas can be exploited to probe thermalization breakdown.

In this work, we show that the DMRG effective Hamiltonian harbors a wealth of spectral information about ergodicity breaking in regimes far beyond previous applications [Fig.~\ref{fig:summary}(c)].
Using the 1D random-field XXZ chain, we find that the effective Hamiltonian captures the thermal-MBL transition, including the level statistics and entanglement structure, in good agreement with previous studies.
We also demonstrate that the same method detects a distinct type of weak ergodicity breaking associated with QMBSs in chaotic models without disorder.
These results establish the DMRG effective Hamiltonian as a versatile tool for exploring far-from-equilibrium physics beyond the reach of ED.

\paragraph{DMRG effective Hamiltonian}
We consider MPS states defined on a finite chain with $N$ sites:
\begin{equation}\label{eq:MPS}
	\ket{\psi(A)}=\sum_{\{\sigma_i\}} A_1^{\sigma_1}A_2^{\sigma_2}\cdots A_N^{\sigma_N}
	\ket{\sigma_1,\sigma_2,\ldots,\sigma_N}\,,
\end{equation}
where $A_i^{\sigma_i}$ are a set of $d$ matrices defined on each site $i$, with bond dimensions
$\chi_{i-1}\times \chi_i$.
We assume an open chain, hence the boundary matrices are vectors with $\chi_0=\chi_N=1$.
We will use the standard graphical representation of $A_i$, where horizontal legs denote the virtual
matrix indices and the vertical leg is the physical index $\sigma\in\{1,2,\ldots,d\}$,
Fig.~\ref{fig:summary}(a).
In the examples below, we will focus on spin-$1/2$ systems with $d=2$.

Importantly, the physical state $\ket{\psi(A)}$ should remain unchanged under a gauge transformation, which can be used to impose left- and right-canonical forms on the MPS~\cite{Schollwock2011}, see the Supplementary Material (SM) for details~\cite{SOM}.
For a bipartition between sites $i$ and $i+1$, which splits the chain into left ($L$) and right ($R$) parts, the canonical forms are related via the so-called center gauge, $A_C^{\sigma_i}=C^{i-1}A_R^{\sigma_i}=A_L^{\sigma_i}C^{i}$, where $C^i$ contains the Schmidt coefficients across the given bipartition~\cite{VidalTEBD}.
In the center gauge, the state can be written in a mixed canonical form
\begin{equation}
	\label{eq:schmidt}
	\ket{\psi(A)}=\sum_{a,b,\sigma_i}[A_C^{\sigma_i}]_{[a,b]}\,
	\ket*{\Psi_{L,a}}\ket{\sigma_i}\ket*{\Psi_{R,b}}\,,
\end{equation}
where $\ket*{\Psi_{L,a}}$ and $\ket*{\Psi_{R,b}}$ form orthonormal Schmidt bases for the
left and right blocks of the state, represented in terms of $A_L^{\sigma_i}$ and $A_R^{\sigma_i}$.

The DMRG algorithm approximates the ground state of a target Hamiltonian by constructing an
effective Hamiltonian, explicitly taking advantage of the mixed canonical form.
Similar to Eq.~(\ref{eq:MPS}), a general Hamiltonian $\Ham$ can be represented as a
matrix-product operator (MPO) with two virtual and two physical indices, Fig.~\ref{fig:summary}(a).
Contracting the MPO representation of $\Ham$ with the MPS everywhere except for $M$ sites yields the
effective Hamiltonian $\Heff$.
By grouping tensor indices, $\Heff$ forms a $\chi_{i-1}\chi_i d^{M}$ dimensional matrix.
While standard DMRG targets the ground state of $\Heff$ as an approximation to the ground state of $\Ham$, we focus instead on the structure of the {\em full} eigenspectrum of $\Heff$, and demonstrate it reveals nonthermal properties of quantum many-body Hamiltonians.

Unless specified otherwise, we take $M=1$ and exactly diagonalize the corresponding $\Heff$, which yields $\chi_{i-1}\chi_i d$ eigenvectors $v_j$ with
energies $E_j$ that can be reshaped into MPS tensors for site $i$, $[v^{\sigma_i}]_{[a,b]}$.
By analogy with Eq.~(\ref{eq:schmidt}), each of these defines an orthogonal state in the full
$d^{N}$-dimensional Hilbert space,
\begin{equation}
	\label{eq:v_states}
	\ket{\psi(A,v_j)}=\sum_{a,b,\sigma_i}[v^{\sigma_i}]_{[a,b]}\,
	\ket*{\Psi_{L,a}}\ket{\sigma_i}\ket*{\Psi_{R,b}}\,,
\end{equation}
where the Schmidt bases $\ket*{\Psi_{L,a}}$ and $\ket*{\Psi_{R,b}}$ are fixed by the
choice of reference MPS tensors $A^{\sigma_n}$ and are therefore identical for all $v_j$.
This construction allows us to compute standard observables, such as the bipartite entanglement entropy, $S_E = -\sum_k \lambda_k^2 \log\lambda_k^2$, where $\lambda_k$ are the Schmidt coefficients for the bipartition of our system immediately to the left of the central site in $\Heff$.

The properties of $\Heff$ depend both on the target Hamiltonian $\Ham$ and on the
 state $\ket{\psi(A)}$.
One may choose $\ket{\psi(A)}$ as an approximate eigenstate of $\Ham$, e.g., obtained using DMRG-S, or as a time-evolved state following a quantum quench.
Below we opt for the former method, which ensures that $\Heff$ provides the best
local approximation to the full quantum Hamiltonian.
We note that, in general, the states $\ket{\psi(A,v_j)}$ are not eigenstates of $\Ham$, reflecting the fact that
$(\Heff)^2\neq(\Ham^2)_\eff$, which we quantify by computing their energy variance~\cite{SOM}.

\begin{figure}[tbp!]
    \includegraphics[width=\columnwidth]{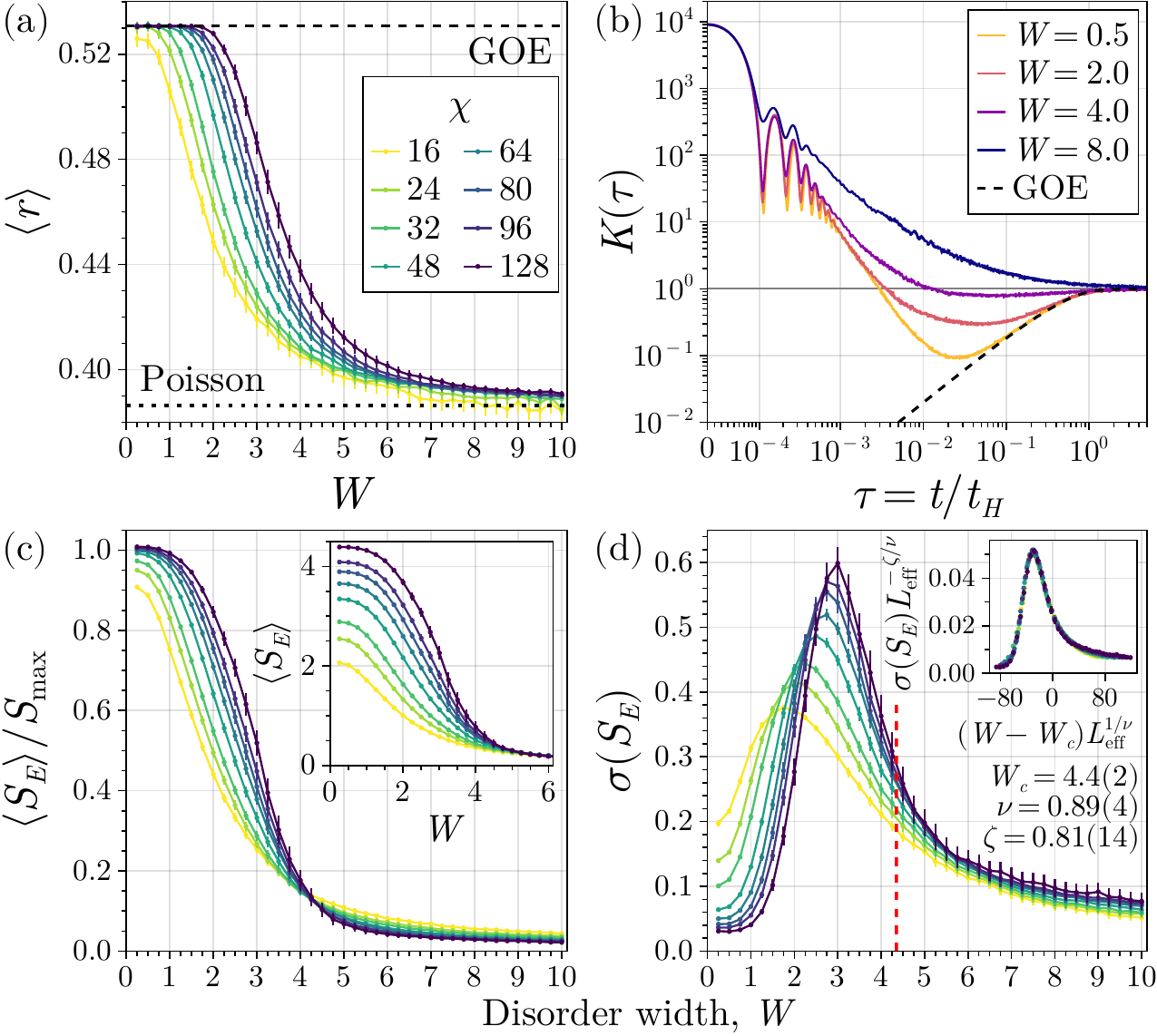}
    \centering
    \caption{%
    	Many-body localization in the effective DMRG Hamiltonian for the random-field XXZ model, Eq.~(\ref{eq:Heisenberg}).
    	(a) The mean level spacing ratio $\langle r \rangle$ as a function of disorder strength $W$ with increasing bond dimension $\chi$.
        The GOE (Poisson) value is marked by a dashed (dotted) line.
        We average over 125--500 disorder realizations and ten central sites; error bars show the 95\% confidence interval.
        (b) The spectral form factor (SFF), $K(\tau)$, against the rescaled time $\tau = t / t_H$, where $t_H$ is the Heisenberg time.
    	The familiar dip-ramp-plateau structure is observed for small $W$ (dashed line shows the GOE prediction), while the dip is absent for sufficiently large disorder.
    	(c) The average entanglement entropy $S_E$, normalized by the maximum entropy (see text), as a function of $W$.
    	The inset shows raw data.
        (d) The standard deviation of $S_E$ between eigenstates in each disorder realization.
        The inset shows a finite-size scaling collapse of the data, along with calculated critical exponents $\nu$,~$\zeta$; the critical disorder $W_c$ (marked by a dashed red line in the main panel); and their 95\% confidence intervals.
		All data are for system size $N = 50$.
    }
    \label{fig:mbl}
\end{figure}

\paragraph{Many-body localization}
As our first example, we study the paradigmatic model of MBL, the 1D spin-$1/2$ XXZ chain in a disordered field:
\begin{equation}\label{eq:Heisenberg}
    \Ham = J \sum_{j=1}^{N - 1} (S^x_j S^x_{j+1} + S^y_j S^y_{j+1} + \Delta S^z_j S^z_{j+1}) + \sum_{j=1}^N h_j S^z_j\ .
\end{equation}
Here, $S_j^\alpha$ are the standard spin-1/2 matrices, the overall energy scale is set to $J=1$, and the $h_j$ are drawn uniformly and independently from the interval $[-W, W]$.
We pick $\Delta = 1$ and $N = 50$ unless otherwise specified.
Note that the model is integrable for $W=0$, while in the non-interacting case ($\Delta = 0$), it exhibits Anderson localization for any disorder strength~\cite{Huse-rev,AbaninRev}.

For each disorder realization and parameter choice, we use the DMRG-S algorithm~\cite{DMRGS} to extract an (approximate) eigenstate $\ket{\Psi_0}$, in MPS form with maximal bond dimension $\chi$, close to energy $E = 0$ (in the middle of the spectrum) and with total magnetization $m^z = \sum_j S^z_j = 0$.
We then form the $M=1$ effective Hamiltonian for this eigenstate and diagonalize it to obtain the eigenvalues $E_i$ and eigenstates $\ket{\psi_i}$, from which we compute  standard diagnostics of quantum chaos.
In addition to disorder averaging, we also average the results for ten sites in the middle of the chain to further suppress fluctuations.
The eigenstates of $\Heff$ may belong to different symmetry sectors from the input MPS, hence they do not necessarily have $m^z = 0$.
We target the largest $m^z$ sector and restrict to the middle-third of the eigenstates of $\Heff$.
The maximal bond dimension $\chi$ (which is saturated in almost all cases) sets an effective length scale $L_\text{eff}$, which is obtained by equating the Hilbert space dimension $d^M \chi^2 = d^{L_\text{eff}}$.

In Fig.~\ref{fig:mbl}(a), we plot the mean of the level spacing ratio,
$r_j = \min(s_j, s_{j+1})/\max(s_j, s_{j+1})$,
where $s_j = E_{j+1} - E_j$ is the spacing between the neighboring energy levels of $\Heff$.
The mean value $\langle r \rangle$ should approach the $r_\text{GOE} \approx 0.536$ for chaotic systems in the Gaussian Orthogonal Ensemble (GOE) class of random matrices, and $r_\text{P} \approx 0.386$ for uncorrelated energy levels with the Poisson statistics~\cite{Oganesyan_Huse_PRB_2007}.
We expect a crossover between these values with increasing $W$, as confirmed in Fig.~\ref{fig:mbl}(a) for sufficiently large $\chi$.
Furthermore, $\langle r \rangle$ attains a plateau at the GOE value at weak disorder, whose width widens with increasing $\chi$.

To probe long-range spectral correlations, we compute the disorder-averaged spectral form factor (SFF),
\begin{equation}\label{eq:sff}
K(\tau)=\bigl\langle \bigl|\sum_{j} e^{-i\epsilon_j \tau}\bigr|^2 \bigr\rangle/Z,
\end{equation}
where $\{\epsilon_j\}$  are the eigenvalues of $H_{\mathrm{eff}}$ after a spectrum unfolding~\cite{Suntajs2020} that sets their density to unity.
The normalization $Z$ ensures that $K(\tau\to\infty)=1$, where time is expressed in units of the Heisenberg time, $\tau=t/t_H$~\cite{SOM}.
In our case, $\chi$ sets an effective length scale, hence the Heisenberg time scales as $t_H \propto d^{L_\text{eff}}$.

The SFF in chaotic systems exhibits a characteristic dip-ramp structure before saturating at unity~\cite{Cotler2017,Das2025}.
$K(\tau)$ for $H_{\mathrm{eff}}$ in Fig.~\ref{fig:mbl}(b) clearly reproduces this behavior at weak disorder.
Increasing $W$ progressively suppresses the dip-ramp feature~\cite{Schiulaz2019}, consistent with the loss of long-range spectral rigidity in the MBL regime~\cite{Prakash2021}.
The onset of the linear ramp in Fig.~\ref{fig:mbl}(b) can be used to extract the Thouless time, which is found to be in good agreement with previous ED estimates~\cite{Suntajs2020,Sierant2020}, see SM~\cite{SOM}.

Finally, we characterize the ergodicity of eigenstates using their entanglement entropy, $S_E$, for the bipartition immediately to the left of the central site in $\Heff$. Since not all eigenstates of $\Heff$ are good approximations to the eigenstates of the full Hamiltonian $\Ham$, we apply a selection procedure based on energy variance to account for this~\cite{SOM}.
In the chaotic regime, mid-spectrum eigenstates are close to random vectors, hence their $S_E$ should scale with $L_\mathrm{eff}$~\cite{Page1993}, with the maximal entropy $S_\text{max} = \log\chi - 1/2$~%
\footnote{For a bipartition with dimensions $(2\chi, \chi)$, the standard Page entropy is $S_\text{Page} \approx \log\chi - 1/4$.
Our data is more consistent with the constant offset $-1/2$, which does not affect the dominant volume-law scaling.}.
Our data at weak disorder in Fig.~\ref{fig:mbl}(c)  are consistent with this volume-law scaling.
On the other hand, in the MBL regime, eigenstates obey an area law, $S_E \propto \text{const.}$, since they can be viewed as product states in the basis of local integrals of motion~\cite{Serbyn13-1,Huse13} and are therefore short-range entangled.
We expect that the mean $S_E$ curves for different system sizes will cross with increasing $W$, as indeed confirmed in Fig.~\ref{fig:mbl}(c).

To estimate the entanglement crossing point in the thermodynamic limit, we apply finite-size scaling to the entropy variance, $\sigma(S_E)$, among different eigenstates, since this quantity is expected to be small in either of the phases while diverging at the transition~\cite{Kjall2014}.
Figure~\ref{fig:mbl}(d) shows a peak in $\sigma(S_E)$ that grows increasingly tall and sharp with $\chi$, with a clear trend towards zero deep into the two phases.
We perform a scaling collapse based on $\sigma(S_E) = L_\mathrm{eff}^{\zeta/\nu} f[(W - W_c) L_\mathrm{eff}^{1/\nu}]$, which yields a critical disorder of $W_c = 4.4 \pm 0.2$, and critical exponents $\nu = 0.89 \pm 0.04$, $\zeta = 0.81 \pm 0.14$ (with 95\% confidence intervals).
Our estimate is consistent with ED studies that locate the critical point in the range $3.7 \lesssim W_c \lesssim 5.5$~\cite{PalHuse, Kjall2014, Luitz2015, Serbyn15, KhemaniEnt}.
We note, however, that estimates based on many-body resonances give much larger $W_c$~\cite{Morningstar2022}, while the possibility of $W_c$ diverging with system size is also  debated~\cite{Suntajs2020,AbaninChallenges,SelsPolkovnikov2021,deroeck2025absencenormalheatconduction,Weisse2025}.

\begin{figure}[tbp!]
	\includegraphics[width=\columnwidth]{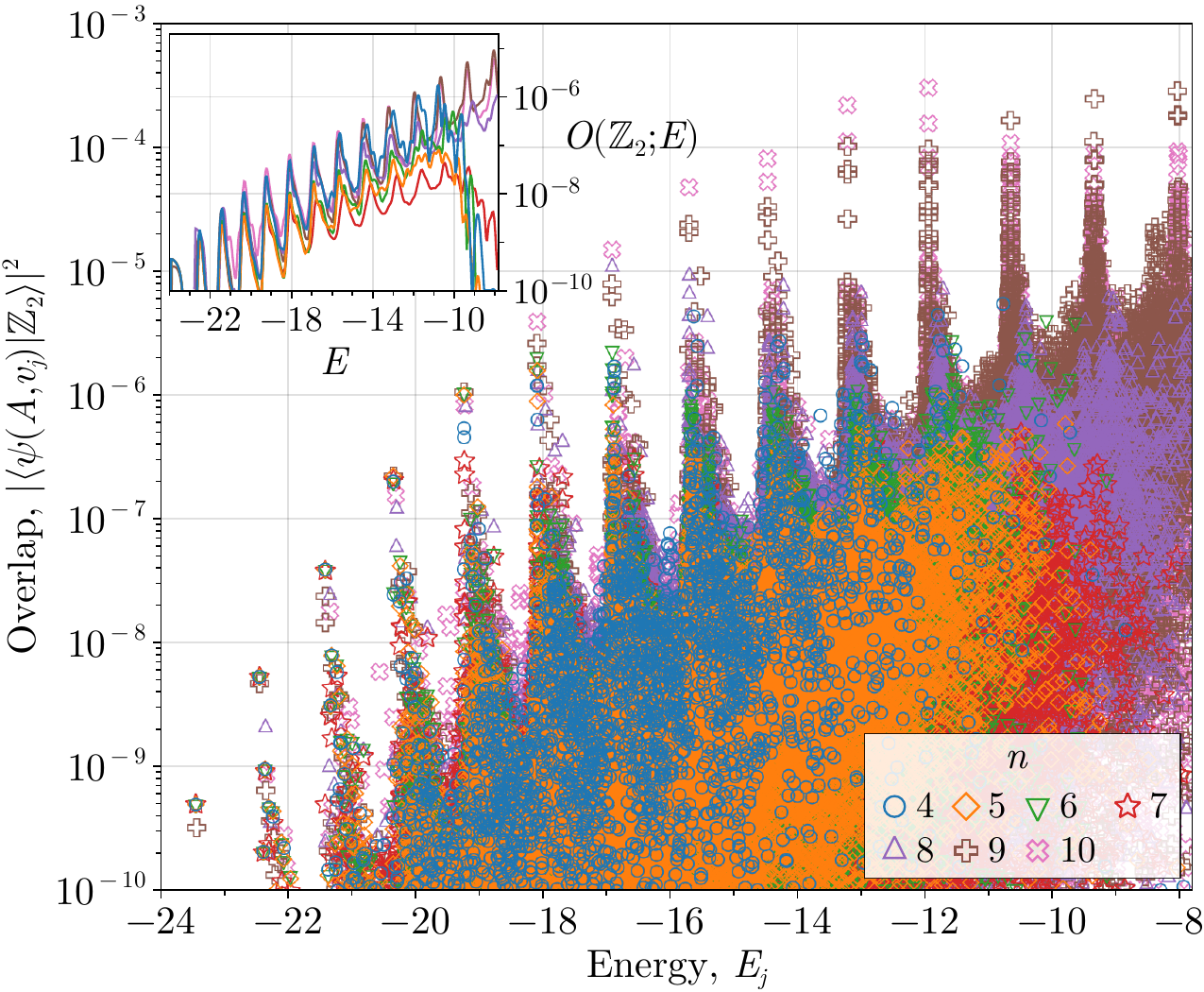}
	\centering
	\caption{%
        Quantum many-body scarring in the effective DMRG Hamiltonian for the PXP model, Eq.~(\ref{eq:PXP}). We target the eigenstate at $E_{\mathrm{target}}(n)=-1.331(N/2-n)$ via DMRG-S and use it to construct $H_{\mathrm{eff}}$. The kinetic constraint is encoded as an additional energy penalty term~\cite{SOM}. The main plot shows the overlaps between the eigenstates of $H_{\mathrm{eff}}$, $|\psi(A,v_i)\rangle$ and the $|\mathbb{Z}_2\rangle$ state for $n\in[4,10]$, while the inset shows the same data convolved with a Gaussian filter $e^{-(E-E_i)^2/\sigma^2}$ with $\sigma=0.1$.
        All results are for $N=40$ and $\chi=150$.
    }
	\label{fig:pxp}
\end{figure}
\paragraph{Quantum many-body scars}%
A distinct form of ergodicity breaking occurs in systems that host non-thermalizing eigenstates known as quantum many-body scars (QMBSs)~\cite{Serbyn2021, MoudgalyaReview, ChandranReview}.
There is a plethora of models and mechanisms giving rise to QMBSs, such as
eigenstate embedding~\cite{ShiraishiMori}, spectrum-generating algebras~\cite{BernevigEnt,Iadecola2019_2,Mark2020,ODea2020Tunnels,Moudgalya2022}, dynamical symmetries~\cite{Buca2023}, group-theoretic constructions~\cite{Pakrouski2020,Ren2020}, and semiclassical limits~\cite{wenwei18TDVPscar,Michailidis2020,Turner2020,Evrard2024,Pizzi2025}.
Here we focus on a paradigmatic PXP model describing 1D Rydberg atom arrays~\cite{FendleySachdev,Turner2017,Surace2020}:
\begin{equation}\label{eq:PXP}
 \mathcal{H}_{\mathrm{PXP}}= \frac{\Omega}{2} \sum_{j=1}^N P_{j-1} S^x_{j} P_{j+1}, \quad P_{j}\!=\!\frac{1}{2}-S_j^z,
\end{equation}
where the Rabi frequency can be set to $\Omega\!=\!1$ and, with open boundary conditions, the first and last terms are understood to be $S_1^x P_2$ and $ P_{N-1} S_N^x$.
The projectors $P$ encode a kinetic constraint: a spin can flip only if both of its nearest neighbors are in the $\downarrow$ state.

The PXP model is chaotic yet exhibits various signatures of ergodicity breaking, including exact ETH-violating eigenstates~\cite{lin2018exact,ivanov2025exactarealawscareigenstates} and superdiffusive energy transport~\cite{Ljubotina2023}.
Most notably, it shows coherent dynamics when quenched from certain initial states, such as $|\mathbb{Z}_2\rangle\!\equiv\! |{\uparrow}{\downarrow}{\uparrow}{\downarrow}{\cdots}\rangle$~\cite{Turner2018b, RenScarFinder}.
The overlap of energy eigenstates of $\mathcal{H}_{\mathrm{PXP}}$ with the $|\mathbb{Z}_2\rangle$ state reveals characteristic tower structures~\cite{Turner2017}, roughly equidistant in energy with spacing $\Delta E\approx1.331$~\cite{Turner2018b}.
These QMBS eigenstates can be viewed as forming a representation of an emergent collective spin, whose coherent precession describes the revival dynamics of the $|\mathbb{Z}_2\rangle$ state~\cite{Choi2018,Iadecola2019,Bull2020,Omiya2022}.
We next demonstrate that these QMBS towers of states can be revealed using the effective DMRG Hamiltonian.

Using DMRG-S, we target eigenstates of the PXP model at energies $E_{\text{target}}(n)=-1.331(N/2-n)$ which respectively approximate the location of the $n$th scar tower, where $n=0$ approximately corresponds to the ground state and $n=N/2$ to the middle of the spectrum.
After forming $\Heff$, we plot the overlap of its eigenstates with the reviving $|\mathbb{Z}_2\rangle$ state in Fig.~\ref{fig:pxp}.
We find that $\Heff$ reproduces the QMBS tower structure of the spectrum around the target energy, and in particular the most non-thermal states with the highest overlap within each tower.
Curiously, after applying a Gaussian filter to the same data [inset of Fig.~\ref{fig:pxp}], we observe that the target QMBS eigenstate at $E_{\text{target}}(n)$ accurately captures all other QMBS towers with lower energies ($n^\prime < n$), whereas the ones with higher energies ($n^\prime > n$) are much more poorly captured (this is further corroborated by the energy variance data in the SM~\cite{SOM}).
We attribute this asymmetry to the approximate su(2) algebraic structure of QMBS towers~\cite{Choi2018,Iadecola2019,Bull2020,Omiya2022}, and in the SM~\cite{SOM} we prove this for a class of QMBS models where such algebra is exact. 

While in Fig.~\ref{fig:pxp} we targeted QMBS towers at energies approximately known from previous studies, for models in which this is not known in advance, one could first reconstruct the spectrum in the vicinity of the ground state and then proceed iteratively to determine 
$E_{\text{target}}(n+1)$ from the effective Hamiltonian constructed at $E_{\text{target}}(n)$.  Furthermore, we note that in many cases the QMBS states can also be identified by their low entanglement compared to the thermal bulk of the spectrum, as sketched in Fig.~\ref{fig:summary}(c), as we show for PXP model in the SM~\cite{SOM}. 

\paragraph{Conclusions and discussion}%
We have demonstrated that the DMRG effective Hamiltonian contains rich spectral information about quantum many-body systems far from their ground state, including both strong- and weak-ergodicity breaking regimes associated with MBL and QMBSs. In the End Matter, we demonstrate that the DMRG effective Hamiltonian is particularly useful as a spatially-resolved probe of ergodic bubbles, allowing the study of the avalanche instability of MBL.
Crucially, all these diagnostics can be performed in system sizes beyond the reach of ED.

We stress that the effective Hamiltonian spectrum should be interpreted as probing the optimal local representation of a many-body Hamiltonian $\mathcal{H}$ within the $\chi$-controlled variational space, rather than a reconstruction of the full spectrum of $\mathcal{H}$.
In the thermal regime, the level spacing is exponentially small in system size and exact eigenstates obey volume-law entanglement, hence they cannot be faithfully represented at fixed bond dimension $\chi$.
Nevertheless, the observed systematic $\chi$-convergence of spectral diagnostics indicates that our method does capture the relevant local correlations even when the global eigenstate approximation becomes imperfect.
It would be desirable to develop a rigorous understanding of the conditions under which spectral information about $\mathcal{H}$ can be inferred from a restricted variational manifold, which mirrors the previous efforts of reconstructing Hamiltonians or conserved quantities from a few eigenstates or local correlation data~\cite{Garrison2018,Qi2019determininglocal,Li2020HamiltonianTomography,Pawlowski2026LocalIntegralsMotion}.

Our approach opens up several interesting directions.
A natural extension is the interplay between QMBS and MBL physics in disordered QMBS models~\cite{MondragonShem2020,Srivatsa2023,Chen2024Inverting}.
Another avenue are quantum circuit models, which can feature both  Floquet MBL~\cite{Ponte15,Lazarides15} and QMBSs~\cite{Mizuta2020, Mukherjee2020, Rozon2022, Giudici2024, Logaric2024}.
Moreover, it would be interesting to apply our method to other types of ergodicity breaking phenomena, such as Stark MBL~\cite{Schulz2019,vanNiewenburg2019}, Hilbert space fragmentation~\cite{Khemani2020, Sala2020, Scherg2020, Adler2024}, and in particular  disorder-free localization~\cite{Smith2017a, brenes2017many, Chanda2020, Giudici_MBL_U1, Jeyaretnam2025, gyawali2025observationdisorderfreelocalizationusing}, which can be studied directly in the thermodynamic limit~\cite{Paredes2005ExploitingQuantumParallelism, Enss2017ManyBodyLocalizationInfiniteChains}.
Our method could be used as a benchmark for complementary approaches based on flow equations~\cite{Thomson2024, liu2026continuousunitarytransformationsusing}.
More broadly, combining effective Hamiltonian spectra with machine-learning techniques that have been used to identify MBL~\cite{vanNieuwenburg2017, Hsu2018, Doggen2018ManybodyLocalizationDelocalization, Huembeli2019, Theveniaut2019} and QMBSs~\cite{Szolda2022, cenedese2024shallowquantumcircuitsrobust, Feng2025} may enable new data-driven explorations of ergodicity breaking.
In this context, recent proposals for variational preparation of excited states~\cite{millar2025imaginarytimespectraltransforms} raise the possibility of extracting related spectral information directly on quantum processors.

\begin{acknowledgments}
\paragraph{Acknowledgments}%
A.H. and Z.P. acknowledge support by the Leverhulme Trust Research Leadership Award RL-2019-015, and EPSRC Grants EP/Z533634/1, UKRI1337. J.J. acknowledges support by the Leverhulme Trust Grant RPG-2024-112.  Computational portions of this work were undertaken on the AIRE HPC system at the University of Leeds. This research was supported in part by grant NSF PHY-2309135 to the Kavli Institute for Theoretical Physics (KITP).
\end{acknowledgments}

\bibliography{bibliography}

\newpage
\cleardoublepage

\onecolumngrid
\vspace{+1cm}
\begin{center}
{\large {\bf End Matter}}
\end{center}
\vspace{+1cm}

\twocolumngrid

\begin{figure}[tbp!]
    \includegraphics[width=\columnwidth]{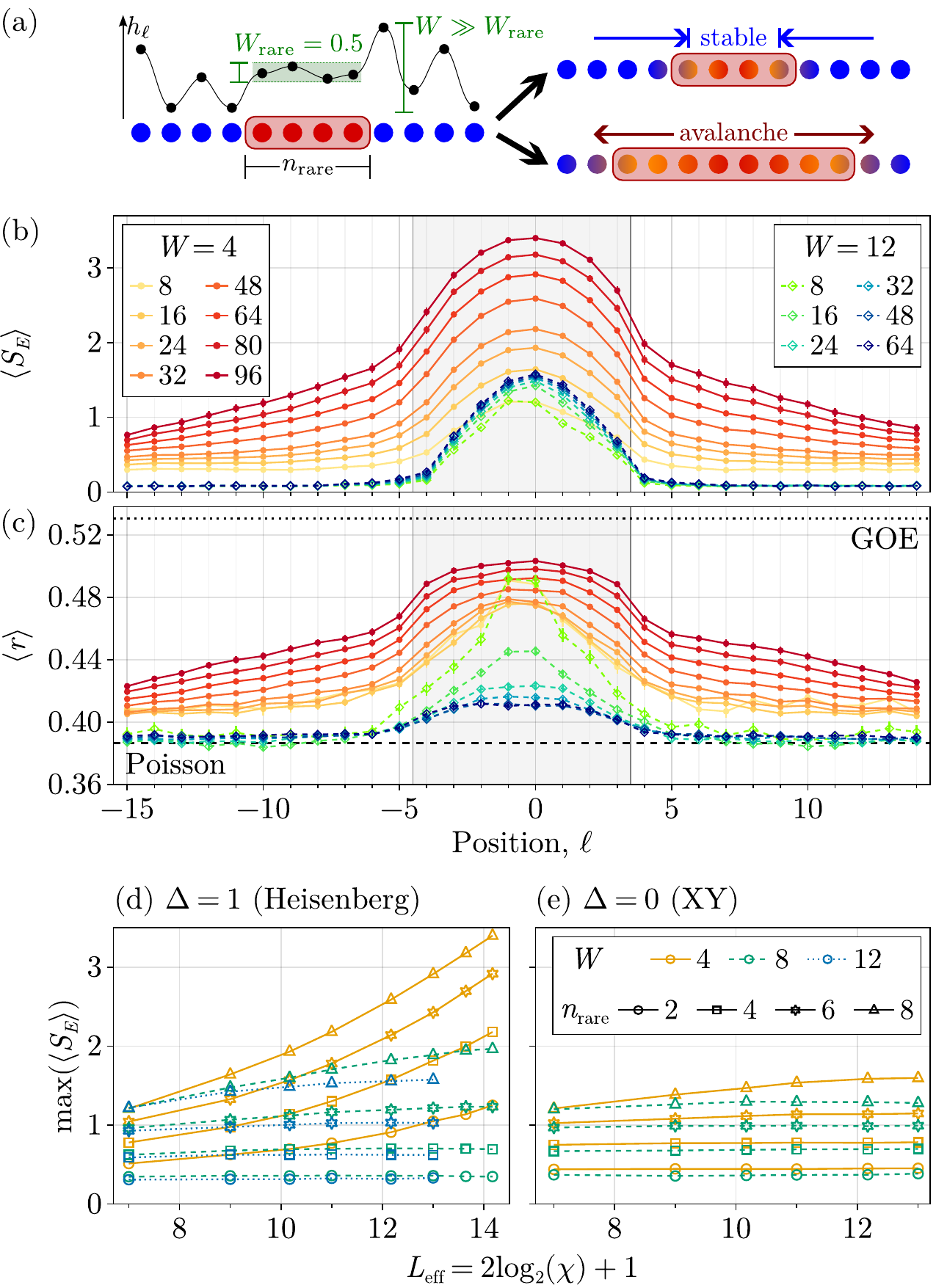}
    \centering
    \caption{%
        (a) A rare low-disorder region forms an ergodic grain that threatens the stability of the MBL phase.
        The thermal grain can grow by thermalizing nearby degrees of freedom: this may arrest, or continue forever in a thermal avalanche.
        We simulate a rare region by setting $W = W_\text{rare} = 0.5$ for $n_\text{rare}$ sites.
        (b) Disorder-averaged bipartite entanglement entropy $\langle S_E \rangle$ in a system with a rare region of size $n_\text{rare} = 8$ (gray shading), with $\Heff$ formed at various offsets $\ell$ from the center.
        We contrast the case of $W=4$ with $W=12$, which are expected to be unstable and stable to avalanches, respectively.
        (c) Same as (b) but for the level spacing ratio $\langle r \rangle$.
        The Poisson (GOE) value is marked with a dashed (dotted) line.
        (d)-(e) The maximum of $\langle S_E \rangle$ with respect to position as a function of $L_\text{eff}$, for various $W$ and $n_\text{rare}$, for (d) the XXZ model $\Delta = 1$, and (e) an XY model with $\Delta = 0$.
        An increase in this quantity suggests susceptibility to an avalanche.
        [In all cases we use at least 500 disorder realizations.]
    }
    \label{fig:rare_regions}
\end{figure}

\paragraph{Ergodic bubbles}%
A central question for the stability of the MBL phase concerns the role of rare thermal inclusions: even deep in the localized regime, spatial fluctuations of disorder can generate locally weakly disordered ``ergodic grains'', which may act as thermal baths for their surroundings.
Such rare regions, if sufficiently large, can trigger an avalanche that destabilizes the MBL phase in the thermodynamic limit~\cite{DeRoeck2017,Gopalakrishnan2019InstabilityManybodyLocalized,Crowley2022}.
We probe this scenario directly within the effective Hamiltonian framework by artificially engineering a rare low-disorder region of size $n_{\mathrm{rare}}$, setting $W = W_{\mathrm{rare}} = 0.5$ on those sites while keeping the surrounding bulk at disorder strength $W$.
This construction mimics a thermal grain embedded in a localized background~\cite{Herviou2019,Colmenarez2024,Sels2023}, as illustrated schematically in Fig.~\ref{fig:rare_regions}(a).
By forming $H_{\mathrm{eff}}$ at varying offsets $\ell$ from the center of the rare region, we obtain spatially resolved diagnostics of how far the thermal influence extends into the bulk.

Figures~\ref{fig:rare_regions}(b) and (c) show the disorder-averaged bipartite entanglement entropy $\langle S_E \rangle$ and mean level spacing ratio $\langle r \rangle$ as a function of position.
For intermediate disorder $W=4$, where the bulk is near the putative transition, the rare region significantly enhances both entanglement and spectral correlations in its vicinity.
The level statistics locally approach GOE behavior and the entanglement entropy develops a broad peak centered on the thermal grain, indicating that the inclusion effectively hybridizes nearby degrees of freedom.
In contrast, for stronger disorder $W=12$, the influence of the rare region remains spatially confined: $\langle r \rangle$ stays close to the Poisson value and $\langle S_E \rangle$ decays rapidly away from the inclusion, consistent with stability against an avalanche.

To quantify this behavior, in Fig.~\ref{fig:rare_regions}(d)-(e) we plot the maximum entanglement entropy (with respect to position) as a function of the effective length $L_{\mathrm{eff}}$.
For the interacting XXZ model ($\Delta = 1$), we observe that at weaker bulk disorder the maximal entanglement grows with $L_{\mathrm{eff}}$, signaling that the rare region increasingly hybridizes distant degrees of freedom and suggesting susceptibility to avalanche instability.
By contrast, at larger disorder the growth is strongly suppressed.
In the noninteracting XY limit ($\Delta = 0$), this enhancement is absent for sufficient $\chi$, consistent with the fact that Anderson localization is stable against such thermal inclusions.

\newpage
\cleardoublepage

\onecolumngrid

\begin{center}
\textbf{\large Supplemental Online Material for ``Ergodicity breaking in matrix-product-state effective Hamiltonians''}\\[5pt]
\vspace{0.1cm}
\begin{quote}
{\small This Supplementary Material provides technical details of the numerical methods and analyses presented in the main text:\linebreak[2]
	\textbf{(i)}~the construction of the DMRG effective Hamiltonian;\linebreak[2]
	\textbf{(ii)}~the implementation of DMRG-S for targeting excited eigenstates;\linebreak[2]
	\textbf{(iii)}~the diagnostics used to assess the quality of effective Hamiltonian eigenstates;\linebreak[2]
	\textbf{(iv)}~data for a single disorder realization of the XXZ model;\linebreak[2]
	\textbf{(v)}~details of the unfolding procedure employed in the spectral form factor calculation, including the extraction of the Thouless time and a finite-size collapse;\linebreak[2]
	\textbf{(vi)}~calculation of the spectral function found in the off-diagonal part of the ETH ansatz;\linebreak[2]
	\textbf{(vii)}~additional data supporting the finite-size scaling analysis for the XXZ model;\linebreak[2]
	\textbf{(viii)}~application to the analytically tractable spin-1 XY model that hosts exact quantum many-body scar eigenstates; \textbf{(ix)}~further results for entanglement entropy and energy variance of quantum many-body scar eigenstates of the PXP model.
}\\[20pt]
\end{quote}
\end{center}

\setcounter{equation}{0}
\setcounter{figure}{0}
\setcounter{table}{0}
\setcounter{page}{1}
\setcounter{section}{0}
\renewcommand{\theequation}{S\arabic{equation}}
\renewcommand{\thefigure}{S\arabic{figure}}
\renewcommand{\thesection}{S\arabic{section}}
\renewcommand{\thepage}{\roman{page}}
\renewcommand{\thetable}{S\arabic{table}}

\makeatletter
\renewcommand{\c@secnumdepth}{0}
\makeatother

\vspace{0cm}

\thispagestyle{empty}

\section{DMRG effective Hamiltonian}

A matrix-product state (MPS) over $N$ lattice sites with open boundary conditions is defined as~\cite{CiracRMP}:
\begin{equation}
	\ket{\psi(A)}=\sum_{\{\sigma_i\}} A_1^{\sigma_1}A_2^{\sigma_2} \cdots A_N^{\sigma_N}\ket{\sigma_1,\sigma_2,\dots,\sigma_N}\,,
\end{equation}
where $A^\sigma_i$ are a set of matrices with bond dimensions $\chi_{i-1}\times \chi_i$.
Each site contains $d$ matrices, $\sigma_i=1,2,\ldots,d$ (we will be mainly interested in spin-1/2 systems, i.e., $d=2$), and the matrices can be site-dependent, hence we distinguish them by index $i=1,2,\ldots,N$.
For an open chain, the boundary matrices must be vectors such that the matrix product evaluates to a scalar for any combination of $\{\sigma_i\}$, hence $\chi_0=\chi_N=1$.

In numerical implementations, the maximum bond dimension is fixed to a particular value, $\chi$, which bounds the maximum bipartite entanglement entropy to $S_E\leq\log(\chi)$ \cite{Schollwock2011}.
The entanglement entropy is defined as the von Neumann entropy $S_E = - \mathrm{tr} \rho_A \log \rho_A$ of a reduced density matrix $\rho_A$ describing a subsystem of the chain containing the left $N_A$ sites.
Thus, the MPS can represent area-law entangled states in 1D, for which $S_E \leq \mathrm{const}$. On the other hand, generic many-body eigenstates are volume-law entangled, $S_E\propto N$, hence they cannot be efficiently represented by an MPS in the thermodynamic limit $N\to\infty$.

An important property of MPS is that the physical state $\ket*{\psi(A)}$ remains unchanged under the gauge transformation $A^\sigma_i\rightarrow A^\sigma_i G$, $A^\sigma_{i+1}\rightarrow G^{-1}A^\sigma_{i+1}$.
This gauge freedom can be used to impose left- and right-canonical forms on MPS that are useful for numerical algorithms:
\begin{equation}
	\label{eq:AL_AR}
\sum_{\sigma_i}A_L^{\sigma_i\dagger}A_L^{\sigma_i}=\mathbb{I}, \quad \sum_{\sigma_i}A_R^{\sigma_i}A_R^{\sigma_i\dagger}=\mathbb{I}.
\end{equation}
These two canonical forms are related via $A^{\sigma_i}_C=C^{i-1}A_R^{\sigma_i}=A_L^{\sigma_i}C^{i}$, where $A^{\sigma_i}_C$ is called the center gauge and
$C^i$ corresponds to the Schmidt coefficients of $\ket{\psi(A)}$ for a bipartition between sites $i$ and $i+1$, which splits the chain into the left subsystem $L$ (containing sites $[1:i]$) and the right subsystem $R$ (sites $[i+1:N]$).
Using the center gauge, the state can be expressed in a mixed canonical form~\cite{VidalTEBD}:
\begin{equation}
	\label{eq:Ac}
	\ket{\psi(A)}=\sum_{a,b,\sigma_i}[A^{\sigma_i}_C]_{[a,b]}\ket{\Psi_{L,a}}\ket{\sigma_i}\ket{\Psi_{R,b}},
\end{equation}
where $\ket*{\Psi_{L,a}}$ and $\ket*{\Psi_{R,b}}$ form orthonormal Schmidt bases for the left and right blocks of the state, respectively.
The two sets of bases are represented in terms of corresponding left and right canonical MPS matrices, $A_L^{\sigma_i}$ and $A_R^{\sigma_i}$.

The numerical algorithms, such as the density-matrix renormalization group (DMRG), directly take advantage of the center-canonical form.
DMRG approximates the ground state of a quantum many-body Hamiltonian by constructing an effective Hamiltonian.
First, a quantum many-body Hamiltonian $\Ham$ is represented as a matrix-product operator (MPO):
\begin{equation}
	\Ham = \sum^d_{\{\sigma_i,\sigma^\prime_i\}=1}
	H^{\sigma_1,\sigma^\prime_1}...H^{\sigma_N,\sigma^\prime_N}
	\ket{\sigma_1,...,\sigma_N}\bra{\sigma^\prime_1,...,\sigma^\prime_N}.
\end{equation}
DMRG works by iteratively sweeping through the system, solving an effective eigenvalue problem for a matrix formed in the manner illustrated in Fig.~1 of the main text, repeatedly finding the lowest energy eigenvector of this matrix (see also Sec.~\ref{app:dmrgs} for details of DMRG-S).
In this process, the MPO representation of the Hamiltonian is contracted with the MPS everywhere in the chain apart from a contiguous region containing $M$ sites.
To the left of the chosen partition, the Hamiltonian is iteratively contracted with $A^{\sigma_i}_L$ and $\bar{A}^{\sigma_i}_L$ to form a rank-$3$ tensor $L(i)$, where $L(0)=1$.
Similarly, to the right the Hamiltonian is iteratively contracted with $A^{\sigma_i}_R$ and $\bar{A}^{\sigma_i}_R$ to form a rank-$3$ tensor $R(i)$, where $R(N+1)=1$.
Finally, the DMRG effective Hamiltonian is a tensor contraction of these two objects with some $M$ uncontracted central sites.
For example, for $M=1$, the explicit expressions are:
\begin{gather}
	\label{eq:L}
	[L(i)]_{b,d,f}=\sum_{a,c,e,\sigma_i,\sigma^\prime_i}[L(i)]_{a,c,e}[A^{\sigma_i}_L]_{a,b}[H^{\sigma_i,\sigma^\prime_i}]_{c,d}[\bar{A}^{\sigma^\prime_i}_L]_{e,f}\ ,\\
	\label{eq:R}
	[R(i)]_{a,c,e}=\sum_{b,d,f,\sigma_i,\sigma^\prime_i}[A^{\sigma_i}_R]_{a,b}[H^{\sigma_i,\sigma^\prime_i}]_{c,d}[\bar{A}^{\sigma^\prime_i}_R]_{e,f}[R(i+1)]_{b,d,f}\ ,\\
	\label{eq:effH}
	[\Heff]_{(a,d,\sigma_i),(c,f,\sigma^\prime_i)}=\sum_{b,e}[L(i-1)]_{a,b,c}[H^{\sigma_i,\sigma^\prime_i}]_{b,e}[R(i+1)]_{d,e,f}\ .
\end{gather}
A graphical representation of the DMRG effective Hamiltonian was shown in Fig.~1 of the main text.
Through a grouping of the $(a,d,\sigma_i)$ and $(c,f,\sigma^\prime_i)$ indices, $\Heff$ can be viewed as a $\chi_{i-1}\chi_id^M$ dimensional matrix.
Typically in DMRG, $M=1$ and $M=2$ are used. Unless specified otherwise, we use $M=1$.

The efficacy of DMRG relies on $\chi$ remaining small.
For gapped, locally interacting 1D Hamiltonians, the ground state and low-energy excited states obey an entanglement area law, hence they can be described using MPS and found with DMRG~\cite{Schollwock2011}.
The DMRG effective Hamiltonian plays a similar role in  MPS algorithms for calculating time-evolution such as the time-dependent variational principle (TDVP)~\cite{haegeman2016}, and algorithms for targeting midspectrum eigenstates such as DMRG-X \cite{Khemani2016,Lim2016,Serbyn16E,Yu2017,Villalonga2018}, or QMBS eigenstates such as DMRG-S \cite{DMRGS}.
Each of these methods uses the effective Hamiltonian to update a vector in the effective vector space, which is subsequently used to update an MPS.

Finally, when the Hamiltonian possesses a global symmetry such as $U(1)$ charge conservation, the MPS can be constrained to be invariant under the symmetry~\cite{CiracRMP}.
The group action of the tensor is simply the tensor-product of the group action on each individual index, leading to the MPS tensor becoming block-sparse.
Consequently, the effective Hamiltonian $\Heff$ becomes block-sparse, although the quantum numbers and dimensions of the irreducible representations are necessarily different from the full quantum many-body Hamiltonian $\Ham$.

\subsection{Properties of effective Hamiltonian}
\label{sec:Properties}

We now discuss the properties of the effective Hamiltonian of Eq.~\eqref{eq:effH}.
The $M=1$ effective Hamiltonian on site $i$ is a $\chi_{i-1}\chi_id$-dimensional matrix, acting upon the center gauge MPS tensor on site $i$, $[A^{\sigma_i}_C]_{[a,b]}$ defined in Eq.~\eqref{eq:Ac}, reshaped into a vector through the grouping of the indices $j=(a,b,\sigma_i)$.
By exactly diagonalizing $\Heff$, we obtain $\chi_{i-1}\chi_id$ eigenvectors $v_j$ with energy $E_j$ which can be reshaped into MPS tensors for site $i$, $[v^{\sigma_i}]_{[a,b]}$.
Each of these corresponds to an orthogonal state in the full $d^N$ dimensional space:
\begin{equation}
	\label{eq:v_states_som}
	\ket{\psi(A,v_j)}=\sum_{a,b,\sigma_i}[v^{\sigma_i}]_{[a,b]}\ket{\Psi_{L,a}}\ket{\sigma_i}\ket{\Psi_{R,b}}\,,
\end{equation}
where $\ket*{\Psi_{L,a}}$, $\ket*{\Psi_{R,b}}$ are orthonormal Schmidt bases for the left and right subsystems.
These are fixed by the initial choice of MPS tensors $A^{\sigma_n}$; therefore, they are identical for every $v_j$.

The properties of $\Heff$ depend upon the full quantum Hamiltonian $\Ham$ but also upon the choice of state $\ket*{\psi(A)}$.
Generally, the states $\ket*{\psi(A,v_j)}$ are not eigenstates of $\Ham$, which can be seen by noting that $(\Heff)^2 \neq (\Ham^2)_\eff$.
The extent to which $\ket*{\psi(A,v_j)}$ approximate physical eigenstates can be quantified through their energy variance,
\begin{equation}\label{eq:energy_variance}
	\sigma_H^2(A,v_j)=v^\dagger_j[(\Ham^2)_{\eff}-(\Heff)^2]v_j =v^\dagger_j(\Ham^2)_{\eff}\,v_j-E^2_j.
\end{equation}
Choosing $\ket{\psi(A)}$ to be an (approximate) eigenstate of $\Ham$ (e.g., by using either DMRG, or DMRG-X/DMRG-S), provides a physically appropriate choice for $\ket*{\Psi_{L,a}}$ and $\ket*{\Psi_{R,b}}$, ensuring that $\Heff$ best approximates the full quantum Hamiltonian.

\section{DMRG-S algorithm}\label{app:dmrgs}
The standard DMRG algorithm \cite{White1992} works by sweeping over an MPS, replacing local tensors by the ground state of the effective Hamiltonian $\Heff$, and thus gradually converging to the ground state of the full Hamiltonian.
This may apply to a single site, forming $\Heff$ with $M=1$, but more commonly this is applied with $M = 2$.
The resulting two-site tensor must then be split into two via a singular value decomposition, with the advantage that this allows the bond dimension to increase (or indeed decrease) during the sweep to capture the desired level of accuracy.
The $M = 2$ algorithm is also less likely to get stuck in local minima.

However, in this work we use the DMRG-S algorithm to obtain excited states of Hamiltonians in MPS format.
DMRG-S \cite{DMRGS} makes a simple modification to the updater: at each step, it applies $(\Heff - \sigma \mathbb{I})^{-2}$ to the local tensor(s), where $\sigma$ is an energy shift, thus iteratively converging to an eigenstate at energy close to $\sigma$.
This can be accomplished efficiently by solving the linear system $(\Heff - \sigma \mathbb{I})^{2} \ket{\phi'} = \ket{\phi}$ with Krylov methods \cite{Haegeman_KrylovKit_2024} for the new tensor $\ket{\phi'}$.
By squaring the shifted energy operator, we make it positive semi-definite, which improves the stability of the algorithm.
Here, we use $M=2$ to improve convergence and allow the bond dimension to grow. Note, however, that we use $M=2$ to obtain the target state via DMRG-S, but for the subsequent analysis of the effective Hamiltonian, as pointed out in the main text, we find it sufficient to use $M=1$.

Finally, an alternative to DMRG-S is the DMRG-X algorithm \cite{Khemani2016,Lim2016,Serbyn16E,Yu2017,Villalonga2018}, which at each step replaces the local tensor(s) by the eigenstate of the effective Hamiltonian that has the highest overlap with the existing tensor(s).
This was designed specifically for MBL systems, which are expected to be highly structured and for which localization ensures that the effective Hamiltonian is a good approximation.
However, this would have biased the algorithm towards more localized states, and performed poorly on the ergodic side of the transition.
For this reason, we chose DMRG-S, which is less biased towards localized states, although there is still a bias towards states with low enough entanglement to be accurately represented by the limited bond dimension.
Despite this, our spectral form factor results (Sec.~\ref{sec:SFF}) and the spectral function results (Sec.~\ref{sec:spectral_function}) demonstrate that we accurately capture the physics of the XXZ model on the ergodic side of the transition.

\section{Variance selection}\label{app:var_weighting}

While we are able to obtain exact eigenstates of the effective Hamiltonian $\Heff$, these are only approximate eigenstates of the true Hamiltonian $\Ham$ and their accuracy may be quantified using the energy variance, Eq.~\eqref{eq:energy_variance}.
In the MBL phase, the DMRG-S algorithm typically finds many eigenstates with low variances, but there are also many high-variance eigenstates.
We believe this to be a consequence of retaining small singular values in the MPS reference state $\ket{\Psi_0}$ which is then used to construct $\Heff$.
However, while including these small singular values and thus increasing the bond dimension produces some of these spurious approximate eigenstates upon diagonalizing $\Heff$, we also obtain many more good eigenstates.

We therefore need a procedure to pick out the good approximations to true eigenstates of $\Ham$, which we do by only including the states with an energy variance $\sigma^2_H$ below a particular threshold for calculations of the bipartite entanglement entropy.
This threshold is calculated as $\max(10^{-7}, 10^3 \times \sigma^2_{H, \text{min}})$, with $\sigma^2_{H, \text{min}}$ the minimum variance across the middle-$1/3$ of eigenstates (by energy) for a particular realization of $\Heff$.
Note that we do \textit{not} perform this selection procedure for the level spacing ratio $\langle r \rangle$: as a function of multiple energy levels, it is not clear how to correctly apply a variance threshold.
Likewise, we do not apply this procedure for the spectral form factor or the spectral function, which are functions of multiple eigenstates.

\section{Single disorder realization}\label{app:single-realisation}
\begin{figure}[h!]
	\centering
	\includegraphics[width=0.8\linewidth]{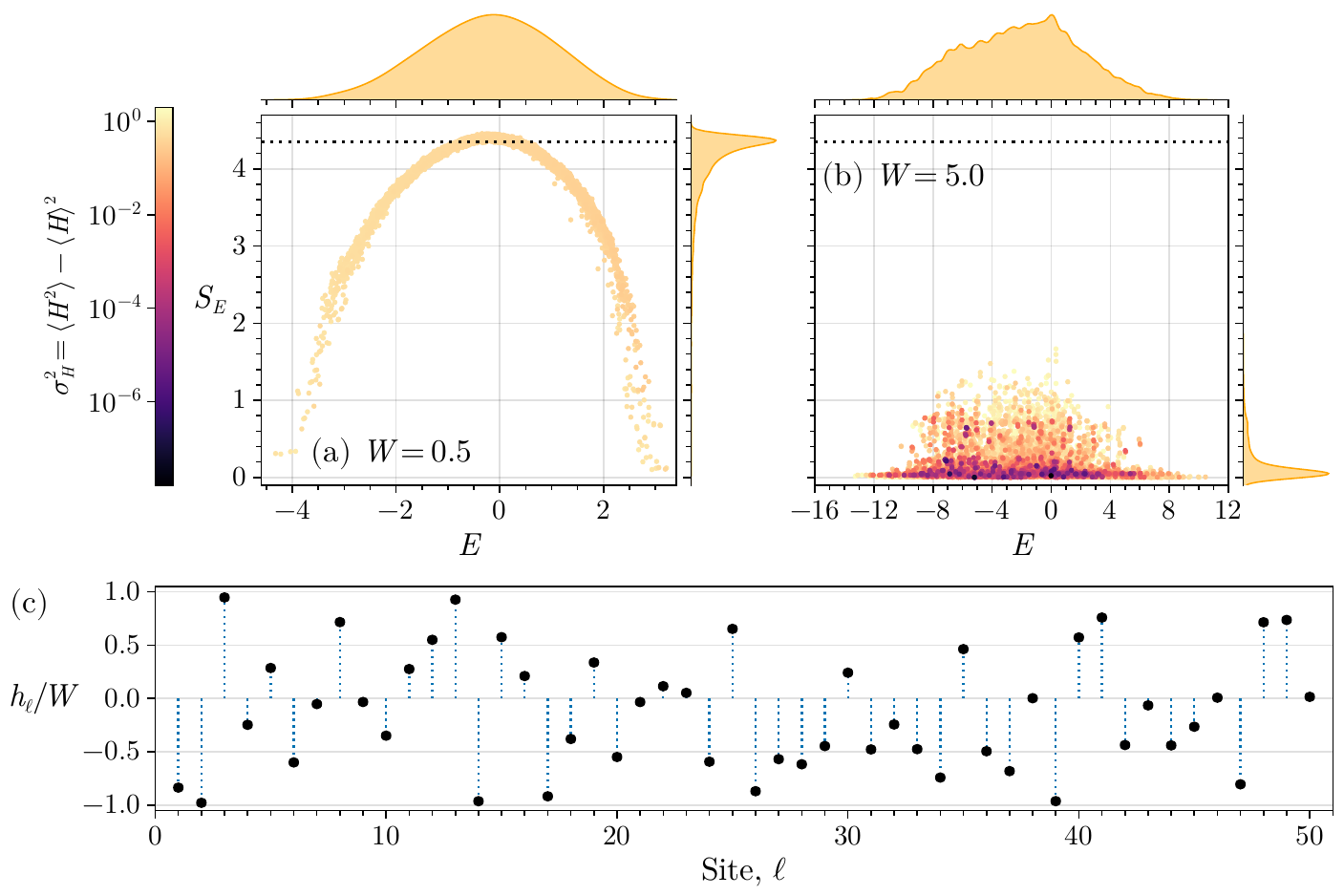}
	\caption{%
		(a)-(b) Bipartite entanglement entropy $S_E$ against energy $E$ for eigenstates of $\Heff$, for a single disorder realization at two values of $W$: (a) $W = 0.5$, and (b) $W = 5$.
		The data are colored according to the energy variance $\sigma^2_H$, with darker colors indicating a lower variance.
		We also show the marginal distributions of $S_E$ and $E$.
		In both cases, we pick site $\ell = 25$ and form $\Heff$ with $M = 1$, then restrict to the $m_z = 0$ sector.
		(c) $h_\ell$ for this disorder realization; see also Eq.~\eqref{eq:single-realisation}.
	}
	\label{fig:single-realisation}
\end{figure}

In Figs.~\ref{fig:single-realisation}(a) and~(b), we show $S_E$ against energy $E$ for the $m^z = 0$ eigenstates of $\Heff$ for a single realization of the XXZ model, with $\chi = 128$ and located in the center of the chain, for $W = 0.5$ and~$5$ respectively.
Specifically, we use the following disorder realization, which is also shown in Fig.~\ref{fig:single-realisation}(c):
\begin{equation}\label{eq:single-realisation}
	\begin{array}{rlllllllll}
		h_\ell = W \times [ -0.836, & -0.978, & +0.948, & -0.247, & +0.286, & -0.600, & -0.052, & +0.717, & -0.033, & -0.349, \\
		+0.276, & +0.550, & +0.927, & -0.963, & +0.575, & +0.211, & -0.917, & -0.380, & +0.337, & -0.548, \\
		-0.033, & +0.116, & +0.053, & -0.593, & +0.653, & -0.869, & -0.569, & -0.617, & -0.446, & +0.242, \\
		-0.478, & -0.243, & -0.475, & -0.741, & +0.463, & -0.495, & -0.681, & +0.003, & -0.963, & +0.573, \\
		+0.759, & -0.436, & -0.064, & -0.440, & -0.265, & +0.008, & -0.805, & +0.715, & +0.737, & +0.016\,]\ .
	\end{array}
\end{equation}
We immediately observe for $W = 0.5$ the smooth curve characteristic of a chaotic system, noting also that the entanglement distribution peaks sharply at the Page value \cite{Page1993}, $S_\text{Page}(\chi) = \log(\chi) - 1/2$.
We also see that the density of states acquires a smooth Gaussian form.
However, the energy variance of these eigenstates \eqref{eq:energy_variance} is relatively large for $W = 0.5$, which reflects the difficulty of obtaining accurate global eigenstates in the delocalized phase.
Despite this, we are still able to reproduce the correct physics in the chaotic regime (see Secs.~\ref{sec:SFF}~\&~\ref{sec:spectral_function}, below).
We can consider this as being due to the thermalizing nature of the chaotic regime: for a system obeying the ETH, local measurements are controlled by the microcanonical ensemble, and vary smoothly from one eigenstate to the next.
It is therefore not an issue if the individual eigenstates obtained are not precise global eigenstates, so long as the left and right bases $\ket*{\Psi_{L,a}}$ and $\ket*{\Psi_{R,b}}$, accessible to $\Heff$ through the bond degrees of freedom, properly capture the physical properties of the model.

On the other hand, for $W = 5$, we see that no such smooth curve emerges, and the average $S_E$ is strongly suppressed compared to $W = 0.5$ -- particularly when we focus on those eigenstates with low energy variance (indicated by darker colors).
The density of states is also highly non-Gaussian and structured, reflecting the change from continuous to discrete local spectrum across the MBL phase transition~\cite{Rahul14}.
These are clear signs of the emergence of localization.

\section{Spectral form factor, unfolding, and Thouless time}\label{sec:SFF}
\begin{figure}[h!]
    \centering
    \includegraphics[width=0.65\linewidth]{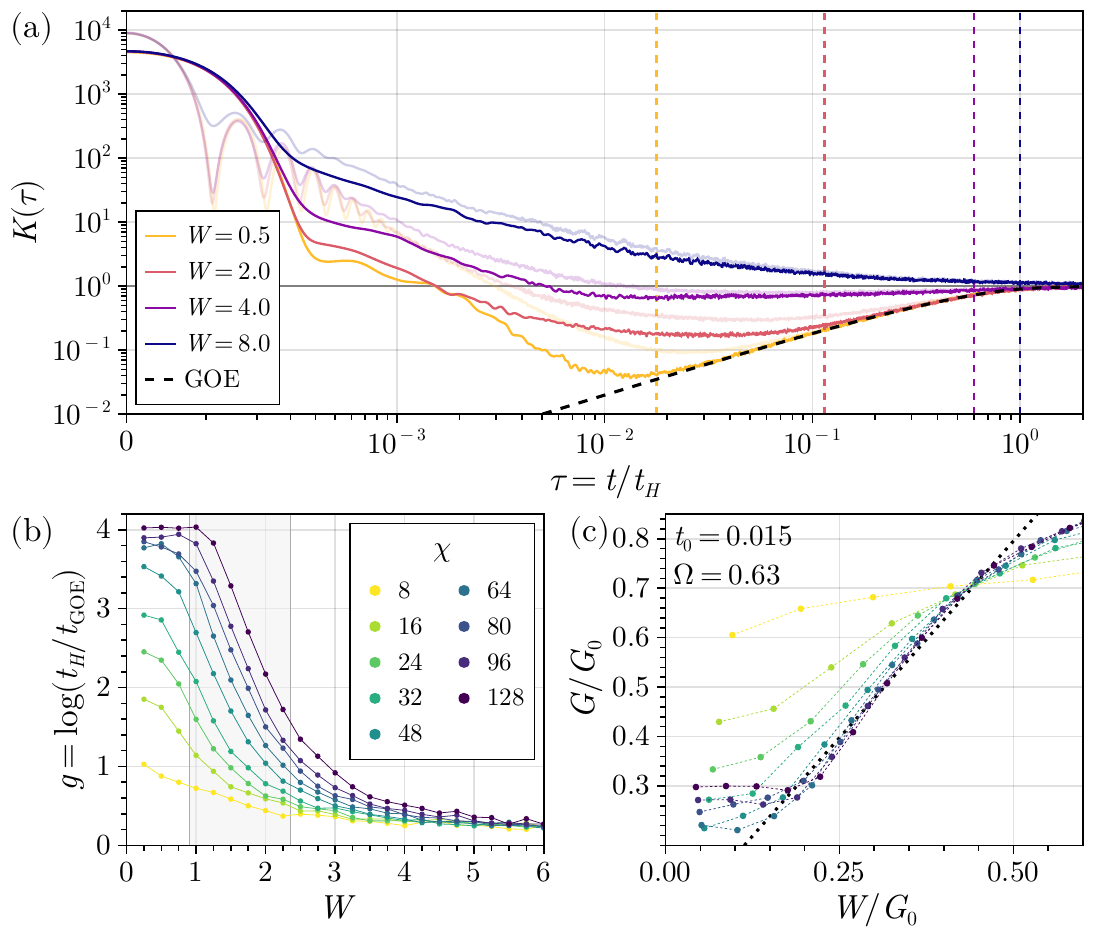}
    \caption{%
        Extraction of the Thouless time from the SFF for the Heisenberg model.
        (a) Filtered SFF for several disorder strengths $W$ (solid lines), with the unfiltered SFF shown for comparison (faded lines), averaged over $125$ disorder realizations for $\chi = 128$.
        The time $\tau_\text{GOE}$ at which the SFF meets the GOE ramp, equivalent to the Thouless time in the chaotic phase, is marked by vertical dotted lines.
        (b) The log-ratio of $t_\text{GOE}$ to the Heisenberg time $t_H$.
        The gray shaded area is used to fit $t_0$ and $\Omega$ (see text).
        (c) Collapse of $G = \log(t_\text{GOE} / t_0 L_\mathrm{eff}^2)$ against $W$ (see text), showing diffusive scaling at small $W$ and large $\chi$.
        The dotted line shows the prediction $G = W / \Omega$.
    }
    \label{fig:sff_thouless}
\end{figure}
In order to compute the spectral form factor (SFF), we must first ``unfold'' the energies $\{E_j\}$ such that the density of states is uniform at large energy scales.
To do this, we follow Ref.~\cite{Suntajs2020} and fit the cumulative distribution function of the energies $D(E) = \#\{E_j < E\}$ to a 10-degree polynomial, $\tilde{D}(E)$, such that the unfolded energies are then given by $\epsilon_j = \tilde{D}(E_j)$.
This procedure removes large-scale fluctuations in the density of states, which are captured by the polynomial fit, while preserving those at small scales.
This further enforces that the Heisenberg time, which is the inverse of the mean level spacing, is $\tau_H = 1$ in rescaled units.
We additionally restrict to those states with $m^z = 0$.

The SFF can then be used to extract the Thouless time $\tau_\text{Th}$, the characteristic timescale for diffusion. In the chaotic regime, this is given by the time $\tau_\text{GOE}$ at which the SFF meets the ``ramp'' of the random matrix prediction for the Gaussian Orthogonal Ensemble (GOE), given by $K_\text{GOE} = 2\tau - \tau\log (1 + 2\tau)$ for $\tau \leq 1$ and $2 - \tau\log ((2\tau + 1)/(2\tau - 1))$ for $\tau > 1$.
However, in the localized regime, the Thouless time is expected to be infinite and $\tau_\text{GOE}$ may cease to be a good estimate~\cite{AbaninChallenges}.
For the extraction of $\tau_\text{GOE}$, we additionally make use of the filtering procedure~\cite{Suntajs2020}:
\begin{equation}\label{eq:sff_filt}
    K_\text{filt.}(\tau)=\biggl\langle \biggl| \sum_{j} f(\epsilon_j) e^{-i\epsilon_j \tau} \biggr|^2 \biggr\rangle /Z\,,\quad
    f(\epsilon) = \exp(-\frac{1}{2} \left[\frac{\epsilon - \mu_\epsilon}{\eta\sigma_\epsilon}\right]^2)\,,
\end{equation}
with $\mu_\epsilon$ and $\sigma_\epsilon$ the mean and standard deviation of the unfolded energies $\{\epsilon_j\}$, and $\eta = 0.5$.
The filtering removes fluctuations at small times, corresponding to the tails of the spectrum and low-energy physics, while prioritizing information from the middle of the spectrum which corresponds to generic behavior.

In Fig.~\ref{fig:sff_thouless}(a) we show the filtered SFF $K_\text{filt.}(\tau)$ for the Heisenberg model with $W = 0.5, 2, 4, 8$, as well as the unfiltered SFF $K(\tau)$, defined in the main text, for reference.
We compute $\tau_\text{GOE}$ as the first time at which $|\log(K_\text{filt.}(\tau) / K_\text{GOE}(\tau))| < 0.15$~\cite{Suntajs2020}, which is marked for each $W$ by vertical dashed lines.
Using $t_H = 1 / \Delta_{E}$, with $\Delta_{E}$ the mean level spacing in the middle of the spectrum, we then obtain $t_\text{GOE} = \tau_\text{GOE} t_H$.
Fig.~\ref{fig:sff_thouless}(b) shows $g = \log(\tau_H / \tau_\mathrm{GOE}) = \log(t_H / t_\mathrm{GOE})$ as a function of $W$.
As expected, this is large and growing with $\chi$ for the chaotic phase, whereas this approaches a small constant for the localized phase.

In the chaotic (and therefore diffusive) regime, we expect $t_\text{Th} \propto L^2$.
Following Ref.~\cite{Suntajs2020}, we fit $t_\text{GOE} = t_0 e^{W / \Omega} L_\text{eff}^2$ in the region $1 \leq W \leq 2.25$ and for $\chi = 128$, to obtain characteristic energy and time scales $\Omega$ and $t_0$.
We then plot $G = \log(t_\text{GOE} / t_0 L_\text{eff}^2)$ against $W$ in Fig.~\ref{fig:sff_thouless}(c), normalizing both axes by $G_0 = \log(t_H / t_0 L_\text{eff}^2)$: here we see a collapse in the chaotic regime (where $t_\text{GOE} \sim t_\text{Th}$), indicating that the Thouless time does indeed scale as $L_\text{eff}^2$ as expected.
This reinforces that we are correctly obtaining the physics of the Heisenberg model, even in the chaotic regime.

\section{Spectral function}\label{sec:spectral_function}
\begin{figure}[tbh!]
	\centering
	\includegraphics[width=0.85\linewidth]{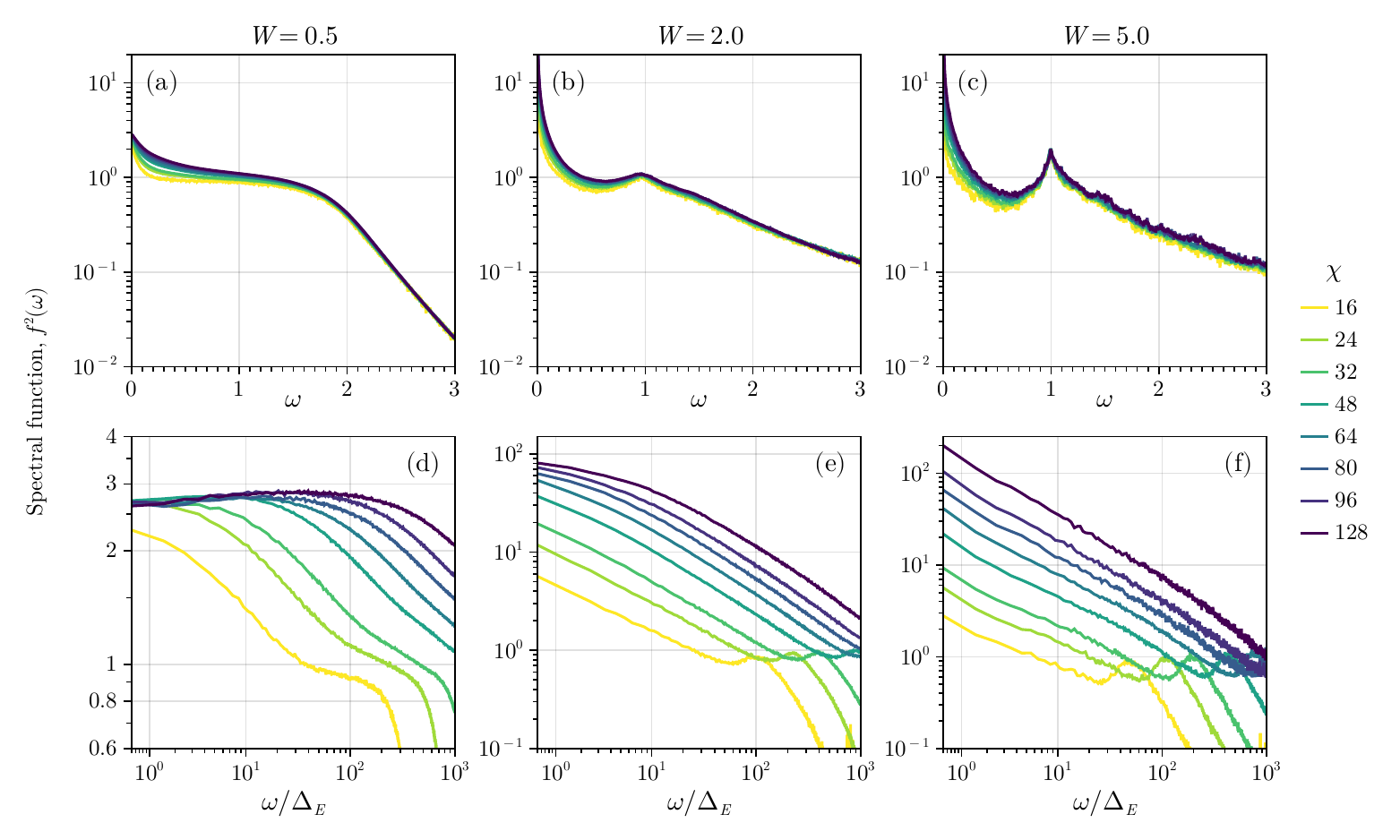}
	\caption{%
		Spectral function for the Heisenberg model, with local operator $S^z_\ell$.
		(a)-(c) The spectral function at large frequencies for $W = 0.5, 2.0, 5.0$, showing the plateau and then power-law decay characteristic of the chaotic phase for $W=0.5$.
		(d)-(f) The spectral function at low frequencies, normalized by the level spacing $\Delta$.
		This data can be compared with Figs.~2~\&~3 of Ref.~\cite{Serbyn-17}, which shows good qualitative and quantitative agreement (see text).
	}
	\label{fig:spectral_function}
\end{figure}
The eigenstate thermalization hypothesis (ETH)~\cite{DeutschETH,SrednickiETH} explains, via the matrix elements of local operators, why isolated chaotic quantum systems approach thermal equilibrium.
The ETH states that, for a local operator $\hat{A}$, the matrix elements between energy eigenstates $\ket{E_j}$ obey the following ansatz,
\begin{equation}\label{eq:eth}
	\matrixel{E_j}{\hat{A}}{E_k} = A(\bar{E})\delta_{jk} + \mathcal{D}^{-1/2} f(\bar{E}, \omega) R_{jk}\,,
\end{equation}
where the average energy $\bar{E} = (E_j + E_k)/2$, the energy difference $\omega = |E_j - E_k|$, $\mathcal{D}$ is the Hilbert space dimension, and $R_{jk}$ is a random complex number, normally distributed with zero mean and unit variance.
The first term ensures that the long-time expectation value of $\hat{A}$ following a quench approaches the microcanonical ensemble value, while the second term guarantees a fast approach to this value and small fluctuations about it thereafter.

In this section, we show that $f(\bar{E}, \omega)$ can successfully be extracted from the DMRG effective Hamiltonian $\Heff$.
We assume that $f(\bar{E}, \omega)$ does not depend strongly on $\bar{E}$ near the middle of the spectrum, in which case we compute~\cite{Serbyn-17}
\begin{equation}
	f^2(\omega) = \mathcal{D} \Bigl\langle \bigl|\!\matrixel{E_j}{\hat{A}}{E_k}\bigr|^2 \delta(\omega - |E_j - E_k|)\Bigr\rangle\,,
\end{equation}
where the average is over both disorder and eigenstates, and we further restrict to $N_\text{avg} = \min(400, \mathcal{D}/20)$ states in the middle of the spectrum.
With this assumption, $f^2(\omega)$ is equivalent to the spectral function, which is measurable in tunneling or absorption experiments~\cite{dAlessio2016}.
The Fourier transform of $f^2(\omega)$ is also closely related to the connected correlation function, $\langle \hat{A}(t) \hat{A}(0) \rangle - \langle \hat A(t)\rangle \langle \hat A(0)\rangle$.

To compute $f^2(\omega)$ in practice, after forming $\Heff$ with $M=1$ and diagonalizing it, we calculate $A_{jk}$ for each $j$ in the window (of width $N_\text{avg}$) and all $k$, binning these by the energy differences $\omega$.
We also bin by the normalized energy differences $\omega / \Delta_E$, where $\Delta_E$ is the mean level spacing within the window (calculated per realization of $\Heff$).
We finally accumulate these bins over both disorder and 10 central sites, then average the values of $|A_{jk}|^2$ within each bin.
Here we do \textit{not} restrict to any particular magnetization sector, although we discard matrix elements that are identically zero due to symmetries.

We show in Fig.~\ref{fig:spectral_function} the results for three different disorder strengths in the Heisenberg model with $\hat{A} = S^z_\ell$ (where $\ell$ is the physical site at which we construct $\Heff$) against both $\omega$ and $\omega / \Delta_E$.
Panels (a-c) may be compared with Fig.~2 of Ref.~\cite{Serbyn-17}, where we see the expected power-law decay at intermediate values of $\omega$, followed by exponential decay at large $\omega$.
We also see that while the curves generally collapse for different $\chi$, they diverge at small $\omega$ in the delocalized phase ($W = 0.5$), a manifestation of the Thouless energy.
Furthermore, we see the same characteristic cusp at $\omega = J$ in the localized phase.

Panels (d-f) may then be compared with panels (a-c) of Ref.~\cite{Serbyn-17}, where we have normalized $\omega$ by the mean level spacing $\Delta$.
Just as for the unnormalized data, we see close quantitative and qualitative agreement.
For small $W$, in the delocalized phase, we see a plateau that grows with increasing $\chi$, since matrix elements do not depend on energy when $\omega$ is below the Thouless energy.
For example, for $\chi = 128$, the plateau indicates that $E_\text{Th} \approx 10^2$, which in turn implies that $\tau_\text{Th} \approx 10^{-2}$, in good agreement with our results in Fig.~\ref{fig:sff_thouless}(a).
We then see the plateau shrink significantly for $W=2.0$ as we approach the localization transition, while for $W = 5.0$ it has clearly shrunk below the level spacing, indicating the onset of MBL.
This data confirms that the DMRG-S effective Hamiltonian correctly reproduces the behavior of local operators on both sides of the localization transition.

\section{Additional finite-size scaling data}\label{sec:fss_additional}
\begin{figure}[htb]
	\centering
	\includegraphics[width=0.85\linewidth]{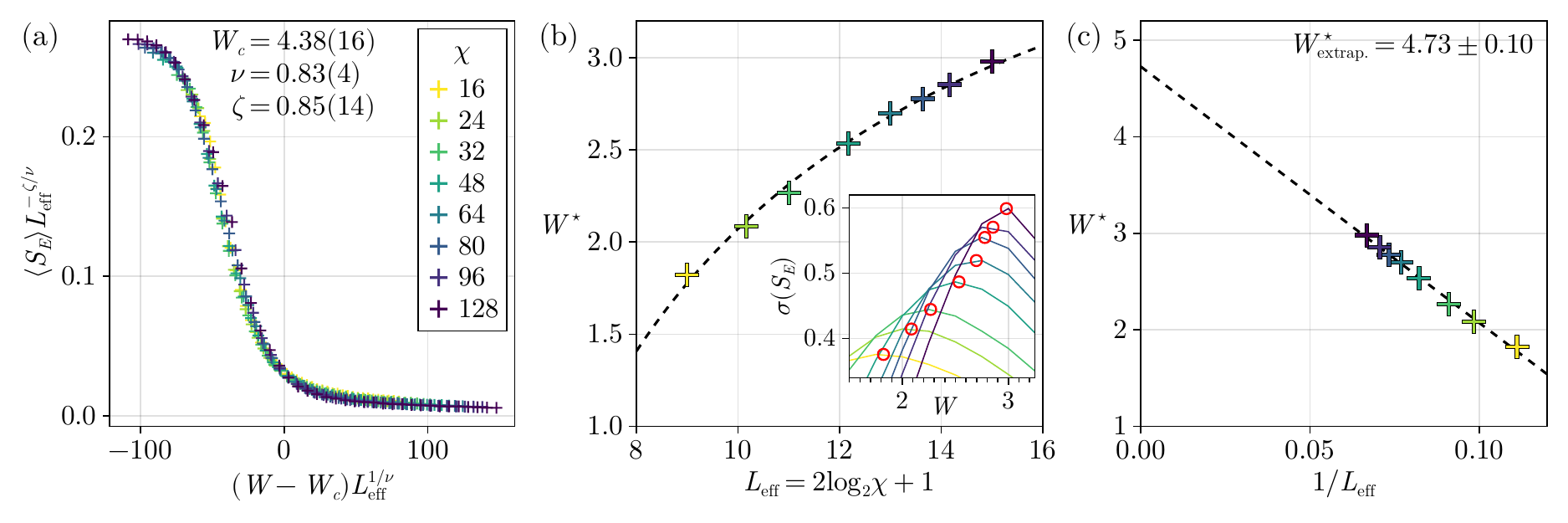}
	\caption{%
		(a) Finite-size scaling collapse for the mean bipartite entanglement entropy, $\langle S \rangle$.
		The obtained critical disorder and critical exponents are also shown, along with their 95\% confidence intervals.
		(b-c) We calculate the location of the peak in the standard deviation of the entanglement entropy, $W^\star$ (red circles, inset), and fit this against $1 / L_\text{eff}$.
		The data (colored crosses) and fit (black dotted line) are shown against both $L_\text{eff}$ and $1 / L_\text{eff}$.
		Extrapolating this to $L_\text{eff} \to \infty$, we obtain $W^\star_\text{extrap.}$, shown with 95\% confidence interval.
	}
	\label{fig:fss_additional}
\end{figure}

In Fig.~2 in the main text, we perform a finite-size scaling collapse of the standard deviation of the eigenstate bipartite entanglement entropy, obtaining the critical disorder $W_c = 4.4 \pm 0.2$, and critical exponents $\nu = 0.89 \pm 0.04$, $\zeta = 0.81 \pm 0.14$ (with 95\% confidence intervals).
We may also perform a collapse of the \textit{mean} of the entanglement entropy, which we show in Fig.~\ref{fig:fss_additional}(a), obtaining $W_c = 4.38 \pm 0.16$, and critical exponents $\nu = 0.83 \pm 0.04$, $\zeta = 0.85 \pm 0.14$ -- consistent with those obtained for the standard deviation to within fitting error.

A different approach is to directly locate the peaks in the standard deviation of the entanglement entropy.
By fitting a quadratic to the points near the maximum for each $\chi$, we estimate the peak location with a resolution finer than the spacing of $W$ in our data; the fitted peak locations $W^\star$ are shown in the inset of Fig.~\ref{fig:fss_additional}(b), along with the original data.
We show this data in Fig.~\ref{fig:fss_additional}(b) and~(c), against both $L_\text{eff}$ and $1 / L_\text{eff}$, where we can see that $W^\star$ lies roughly on a straight line against $1 / L_\text{eff}$.
The intercept of this line with $1 / L_\text{eff} = 0$, corresponding to the limit $L_\text{eff} \to \infty$, gives an alternative estimate of the critical disorder $W^\star_\text{extrap.} = 4.74 \pm 0.10$ (95\% confidence intervals).
This is slightly larger than the estimate given by finite-size scaling collapses.

\begin{figure}[htb]
	\centering
	\includegraphics[width=0.75\linewidth]{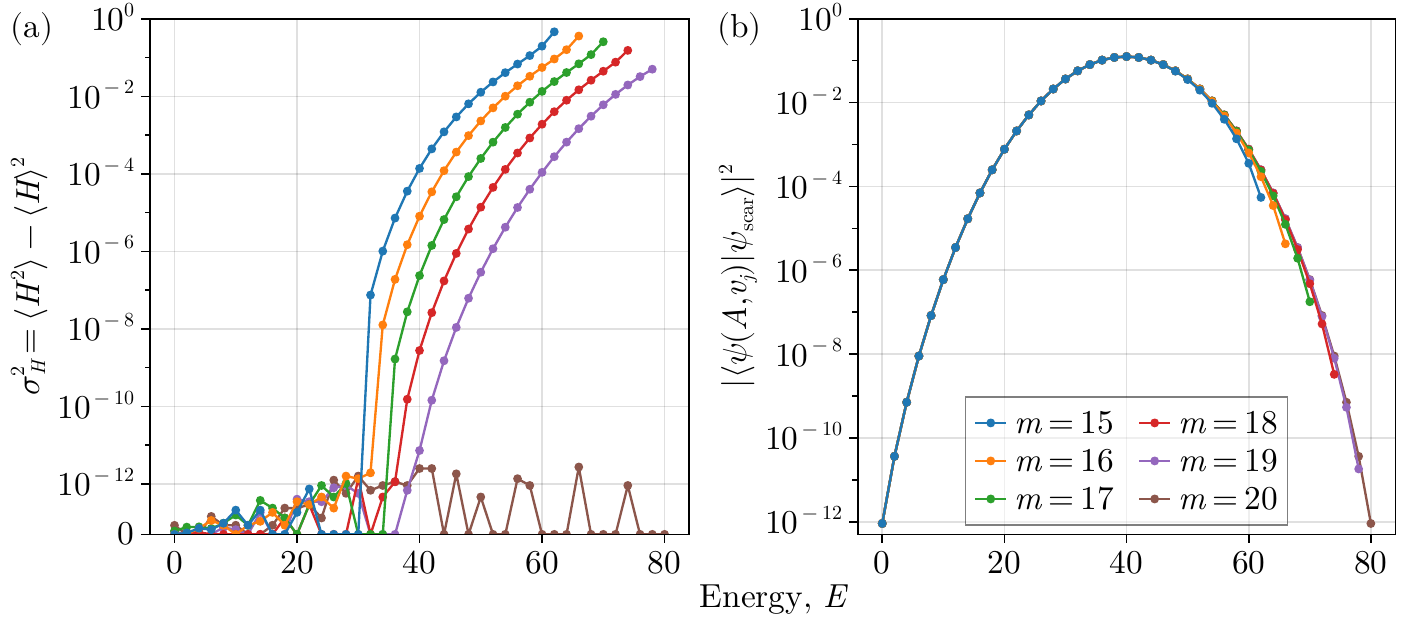}
	\caption{%
		 Quantum many-body scarring in the effective DMRG Hamiltonian $H_{\mathrm{eff}}$ for the Spin-1 XY model, Eq.~(\ref{eq:spin1XY_ham}). (a) The variance of the approximate QMBS eigenstates in the effective Hamiltonian for $m\in[15,20]$. (b) The overlap with the state $|\psi_s\rangle=\bigotimes_j|-1, +1\rangle_{j,j+1}$ for the approximate QMBS eigenstates in the effective Hamiltonian. All results are for $N=40$, $J=1$, $\Delta=1$, $h=1$.
	}
	\label{fig:spin1_xy}
\end{figure}

\section{Effective DMRG Hamiltonian for exact quantum many-body scar models}\label{sec:exactQMBS}

Here we illustrate the effective DMRG Hamiltonian approach on a class of quantum many-body scar (QMBS) models which realize a `restricted spectrum-generating algebra' (RSGA)~\cite{BernevigEnt,Iadecola2019_2,Mark2020,ODea2020Tunnels,Moudgalya2022}. For such models, \emph{exact} QMBS eigenstates are generated by a type of ladder operator, which allows us to analytically understand the underlying principle of the effective DMRG Hamiltonian applied to such models. The PXP model studied in the main text can be viewed as a particular example where the RSGA is inexact~\cite{Choi2018,Iadecola2019,Bull2020,Omiya2022}, but still approximately describes the QMBS eigenstates.

Consider an $N$-site 1D spin chain with arbitrary local Hilbert space dimension $d$ that hosts a tower of exact QMBS.
We assume that the lowest state in the tower is a product state with the following Schmidt decomposition across a bipartition into left and right halves:
\begin{equation}
\begin{split}
\ket{\psi_0} =\bigotimes_{n=1}^{N}\ket{\phi_n} =\left(\bigotimes_{n=1}^{N/2}\ket{\phi_n}\right)\otimes\left(\bigotimes_{n=N/2+1}^{N}\ket{\phi_n}\right) \equiv \ket{\Phi_L}\otimes\ket{\Phi_R}.
\end{split}
\end{equation}
Further, assume that the tower of QMBSs is generated by a local operator of the form
\begin{equation}
Q^+ = \sum_{n=1}^{N} \hat{q}_n = \sum_{n=1}^{N/2} \hat{q}_n + \sum_{n=N/2+1}^{N} \hat{q}_n \equiv Q^+_L + Q^+_R,
\end{equation}
where the local operators satisfy
\begin{equation}
\label{eq:conditions_tower}
\begin{split}
\matrixel{\phi_n}{\hat{q}_n}{\phi_n} &= 0, \\
\hat{q}_n^2 \ket{\phi_n} &= 0.
\end{split}
\end{equation}
The $m$-th scar state is then constructed as
\begin{equation}
\ket{\psi_m} =
\frac{1}{\mathcal{N}(m)} (Q^+)^m \ket{\psi_0},
\end{equation}
where $m=1,2,3,\ldots, N$ is an integer, and $\mathcal{N}(m)$ denotes the norm of the state. Introducing the notation
\begin{equation}
\ket{\psi^{L/R}_m} =
\frac{1}{\mathcal{N}_{L/R}(m)} (Q^+_{L/R})^m \ket{\psi_0},
\end{equation}
the conditions in Eq.~\eqref{eq:conditions_tower} ensure that these vectors are mutually orthogonal.
A straightforward calculation shows that, for $m \le N/2$, the $m$-th QMBS can be written as
\begin{equation}
\ket{\psi_m} =
\sum_{i=0}^{m}
\frac{\mathcal{N}_L(i)\mathcal{N}_R(m-i)}{\mathcal{N}(m)}
\ket{\psi^i_L}\otimes\ket{\psi^{m-i}_R}.
\end{equation}
It follows that all Schmidt vectors appearing in the decomposition of $\ket{\psi_{m_1}}$ are contained within those of $\ket{\psi_{m_2}}$ whenever $0 \le m_1 < m_2 \le N/2$.
A similar argument applies in the complementary regime $N \ge m_1 > m_2 \ge N/2$.

An important example of the above construction is provided by the spin-1 XY model~\cite{Iadecola2019_2},
\begin{equation}
\label{eq:spin1XY_ham}
\mathcal{H}= J \sum_j
\left(S^x_j S^x_{j+1}
+
S^y_j S^y_{j+1}\right)
+
h \sum_j S^z_j
+
\Delta \sum_j (S^z_j)^2,
\end{equation}
where $S^\alpha_j$ are the standard spin-1 operators on site $j$.
This model hosts several towers of exact QMBS at energy $E_n=h(2n-N)+\Delta N$ with the reference state $\ket{\psi_0} = \bigotimes_j \ket{-}_j$ and raising operator $Q^+ = \sum_j (-1)^j (S^+_j)^2$, which fits the above general framework.

To illustrate this approach, we compute the effective Hamiltonian for the $m$th QMBS of the spin-1 XY model for $m\in[15,20]$ at $N=40$, where the resulting effective Hamiltonian provides an approximate representation of all $N+1$ QMBS states.
In Fig.~\ref{fig:spin1_xy}(a), we show the energy variance of these approximate QMBSs.
In agreement with the discussion above, the first $m+1$ QMBSs are exact eigenstates of the Hamiltonian (to machine precision), whereas the remaining $N-m$ states are only approximate eigenstates and exhibit a rapid growth in variance with increasing energy.
The special case is the middle state, $m=N/2=20$, for which the effective Hamiltonian exactly reproduces the full QMBS tower.
In Fig.~\ref{fig:spin1_xy}(b), we plot the overlap with $|\psi_s\rangle=\bigotimes_j|-1, +1\rangle_{j,j+1}$, a product state whose support lies entirely within the QMBS subspace.
We find that the effective Hamiltonian reproduces this overlap profile remarkably well, even in regimes where the variance becomes large, demonstrating that it provides a useful method for capturing the nonthermal behavior of systems hosting QMBSs.
This analytically tractable example of the spin-1 XY model provides a useful starting point for interpreting the results for the PXP model in the main text and in the following section, where an approximate picture still holds, i.e., targeting a QMBS eigenstate at a given energy allows one to approximately reconstruct the QMBS towers at lower energies.

\begin{figure}[htb]
	\centering
	\includegraphics[width=0.75\linewidth]{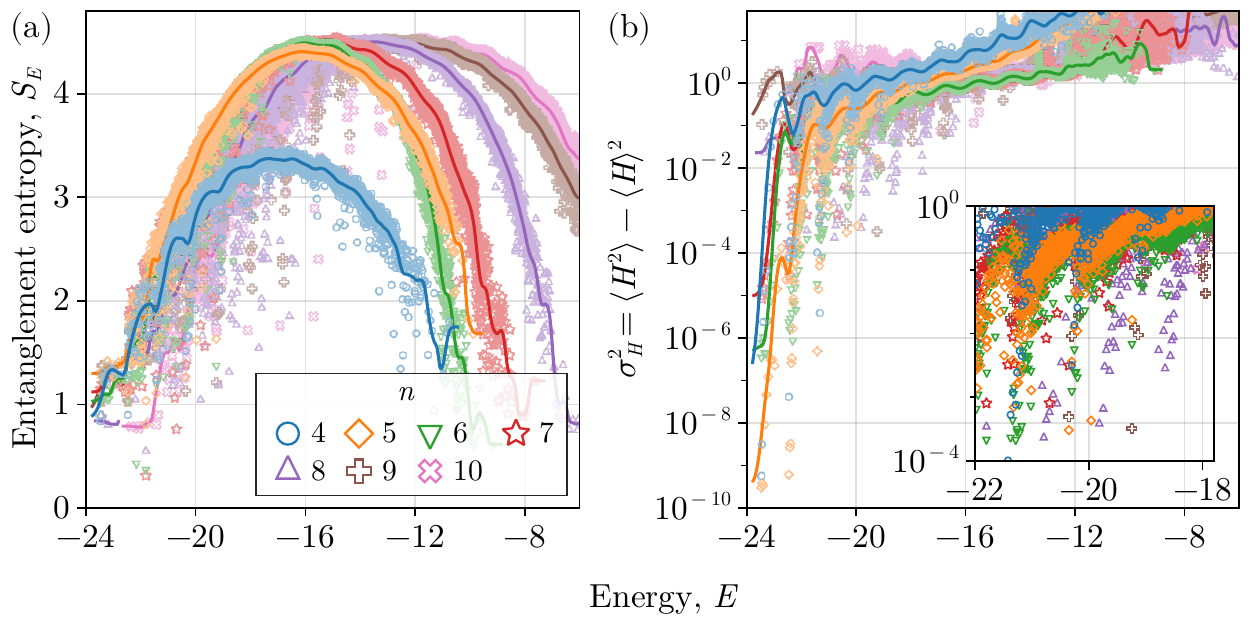}
	\caption{%
		Further results for the quantum many-body scarring in the effective DMRG Hamiltonian for the PXP model, Eq.~(\ref{eq:PXP}).
        Using DMRG-S, we target the eigenstate at $E_{\mathrm{target}}(n)=-1.331(N/2-n)$ and use it to construct $H_{\mathrm{eff}}$. (a) Bipartite entanglement entropy $S_E$ and (b) energy variance $\sigma^2_H = \langle H^2 \rangle - \langle H \rangle^2$, both plotted against energy.
		In each case, we show both individual data points and the smoothed average after applying a Gaussian filter $e^{-(E-E_i)^2/\sigma^2}$ with $\sigma=0.1$.
		For (b) the inset magnifies the region of interest, showing groups of states with low variance. All data are for $N = 40$ and $\chi = 150$.
	}
	\label{fig:pxp_additional}
\end{figure}
\section{Additional data for the PXP model}
In this section, we present additional results for the effective Hamiltonian of the PXP model, obtained by targeting eigenstates at energies $E_{\mathrm{target}}(n)=-1.331(N/2-n)$.
As mentioned in the main text, the PXP model obeys a kinetic constraint which forbids neighboring sites from simultaneously being in the $\uparrow$ state.
For convenience, in our DMRG-S calculations, we enforce this constraint numerically by adding the energy-penalty term, $H_{\text{penalty}}=V_c\sum_j Q_jQ_{j+1}$, where $Q_j=1_j-P_j$. For all results presented in this manuscript, we set $V_c=100$.

As sketched in Fig.~1 of the main text, we expect the PXP model to display broadly ETH-like behavior, ``punctuated'' by a small number of anomalous states with unusually low bipartite entanglement entropy.
This is demonstrated in Fig.~\ref{fig:pxp_additional}(a), where we show the entanglement entropy of eigenstates of $H_{\mathrm{eff}}$ for $n\in[4,10]$.
As expected, the bulk of $H_{\mathrm{eff}}$ spectrum follows the characteristic entropy distribution of an ETH Hamiltonian, albeit within the restricted energy window set by $E_{\mathrm{target}}(n)$ and with an upper bound of $\log\chi=\log150\approx5$.
At the same time, a significant number of states clearly deviate from this ETH-like distribution by exhibiting substantially lower entanglement.
These anomalous states can be identified with QMBSs according to their energies and enhanced overlap with $|\mathbb{Z}_2\rangle$, consistent with the results in Fig.~3 of the main text.
A useful point of comparison is provided by Fig.~\ref{fig:single-realisation} for the non-scarred XXZ model, where we observe clear ETH-like behavior without any anomalous eigenstates of the effective Hamiltonian.

In Fig.~\ref{fig:pxp_additional}(b), we show the energy variance of the eigenstates of $H_{\mathrm{eff}}$ for $n\in[4,10]$.
Once again, we find a set of outlying effective eigenstates with low variance, which correspond to the QMBSs with large overlap with $|\mathbb{Z}_2\rangle$.
More generally, however, the variance of the effective eigenstates is relatively large, and only a small fraction of states satisfy $\sigma_H^2<10^{-2}$.
Nevertheless, as discussed in Sec.~\ref{sec:exactQMBS} and illustrated in Fig.~\ref{fig:spin1_xy}, even when the variance is large, the effective eigenstates can still faithfully capture broader qualitative features of the spectrum.

\end{document}